\documentclass[twocolumn]{aastex631}
\usepackage{times}
\usepackage{amsmath}
\usepackage{graphicx}
\usepackage{color}
\usepackage{array}
\usepackage{multirow}
\usepackage{amssymb}
\usepackage{color}
\usepackage{CJK}

\newcommand{\beq}{\begin{equation}}
\newcommand{\eeq}{\end{equation}}

\newcommand{\delete}[1]{}

\begin{document}

\begin{CJK*}{UTF8}{gbsn}
\title{The spectra of a radiative reprocessing outflow model for fast blue optical transients}


\author{Chun Chen (陈春)}
\affiliation{School of Physics and Astronomy, Sun Yat-sen University, Zhuhai 519082, China}
\affiliation{CSST Science Center for the Guangdong-Hong Kong-Macau Greater Bay Area, Sun Yat-sen University, Zhuhai 519082, China}
\email{chench386@mail2.sysu.edu.cn}

\author[0000-0001-5012-2362]{Rong-Feng Shen (申荣锋)}
\affiliation{School of Physics and Astronomy, Sun Yat-sen University, Zhuhai 519082, China}
\affiliation{CSST Science Center for the Guangdong-Hong Kong-Macau Greater Bay Area, Sun Yat-sen University, Zhuhai 519082, China}
\email{shenrf3@mail.sysu.edu.cn}



\begin{abstract}
The radiation reprocessing model, in which an optically-thick outflow absorbs the high-energy emission from a central source and re-emits in longer wavelengths, has been frequently invoked to explain some optically bright transients, such as fast blue optical transients (FBOTs) whose progenitor and explosion mechanism are still unknown. Previous studies on this model did not take into account the frequency dependence of the opacity. We study the radiative reprocessing and calculate the UV-optical-NIR band spectra from a spherical outflow composed of pure hydrogen gas, for a time-dependent outflowing mass rate. Electron scattering and frequency-dependent bound-free, free-free opacities are considered. The spectrum deviates from the blackbody at NIR and UV frequencies; in particular, it has $\nu L_{\nu} \propto \nu^{1.5}$ at NIR frequencies, because at these frequencies the absorption optical depth from the outflow's outer edge to the so-called photon trapping radius is large and is frequency dependent. We apply our model to the proto-type FBOT AT2018cow by {fitting} the spectra to the observed SED. The best-fit mass loss rate suggests that the total outflow mass in AT2018cow is $M_{\rm out} \approx 5.7^{+0.4}_{-0.4} \, M_{\odot}$. If that equals the total mass lost during an explosion, and if the progenitor is a blue supergiant (with a pre-explosion mass of $\sim 20 \, M_{\odot}$), then it will suggest that the central compact remnant mass is at least $\approx \, \rm{14 \, M_{\odot}}$. This would imply that the central remnant is a black hole. 
\end{abstract}



\section{Introduction} \label{Introduction}

Fast Blue Optical Transients (FBOTs) are a new class of astrophysical transients that have recently been discovered in some optical surveys \citep[]{Drout2014}. They are characterized by their blue colors ($g - r < -0.2$), fast rise (usually $< 10 $ days) and relatively quick decline (usually $> 0.15$  mag/day) in the optical and UV bands, with peak luminosities $L_{\rm peak} > 10^{43}$ erg/s. The physical origin of FBOTs is still unknown due to the extremely rare observations. 

{In recent years, the advancement of multi-wavelength observational facilities has enabled multi-wavelength monitoring of FBOTs. These include CSS161010 \citep[]{Coppejans2020}, ZTF18abvkwla \citep[]{Ho2020}, AT2018cow \citep[]{Prentice2018, Margutti2019, Perley2019, Ho2019}, AT2020xnd \citep[]{Perley2021, Ho2022}, AT2020mrf \citep[]{Yao2022}, AT2022tsd \citep[]{Ho2023}, AT2023fnh \citep[]{Chrimes2024}. These FBOTs exhibit remarkably bright radio emission (typically $L_{\nu \gtrsim 100 \, \rm{GHz}} > 10^{28} \, \rm{erg \, s^{-1} \, Hz^{-1}}$). Additionally, they show luminous X-ray emission (typically $L_X \simeq 10^{43} \, \rm{erg \, s^{-1}}$), lasting from tens of days (e.g., AT2020xnd) to hundreds of days (e.g., CSS161010, AT2020mrf) or even thousands of days (e.g., AT2018cow). The persistent and luminous X-ray emission suggests that FBOTs are likely powered by a central engine. \citep[]{Margutti2019, Yao2022, Chrimes2024}.} 

{Note that the FBOTs discussed in \cite{Drout2014} were selected from existing archival data and lacked multi-wavelength observations. Unlike other well-studied transients (e.g., SNe or GRBs), most FBOTs still suffer from sparse temporal sampling in both photometry and spectroscopy.} Among them, AT2018cow is not only the closest but also exhibits a plethora of multi-wavelength observational data, from X-ray to radio wavelengths \citep[]{Prentice2018, Margutti2019, Perley2019, Ho2019}. {Notably, AT2018cow exhibits a significant NIR excess in the observed spectral energy distribution (SED), suggesting that the high-energy radiation from the central engines is likely reprocessed into the NIR band by optically thick material \citep[]{Margutti2019}. However, due to the lack of well-sampled NIR monitoring in other FBOTs, it remains unclear whether such reprocessing occurs in other FBOTs.}

Nevertheless, the progenitor of AT2018cow {still remains a puzzle}. The typical supernova models fail to explain the light curve features of AT2018cow: the high peak luminosity ($L_{\rm opt, peak} \sim 10^{44} \, \rm{erg \, s^{-1}}$) requires a nickel mass $M_{^{56}Ni} > 5 M_{\odot}$, while the extremely short rising time ($t_{\rm 1/2, rise} \sim 3 \, \rm{days}$) imposes a constraint of the total ejecta mass $M_{\rm ejecta} < 0.01 M_{\odot}$, which is clearly unreasonable \citep[]{Perley2019}.

The observations of AT2018cow also exclude any other known progenitors. The location of AT2018cow is far from the center of its host galaxy, ruling out the scenario of tidal disruption events (TDEs) by a supermassive black hole (BH) \citep[]{Margutti2019}. The presence of a dense CSM environment ($n_{\rm CSM} \approx 9 \times 10^{6} \, \rm{cm^{-3}}$), indicated by the exceptionally bright  ($L_{\nu = 8.5 \rm{GHz}, \rm{peak}} \sim 4 \times 10^{28} \, \rm{erg \, s^{-1} \, Hz^{-1}}$) and prolonged ($\sim \rm{a \, few} \times \, 100 \, days$) radio emission of AT2018cow, effectively excludes the scenarios of AT2018cow being a result of compact binary mergers or {a} TDE by an intermediate-mass BH \citep[]{Yu2013, Metzger2014_2, Chen2018}. {\cite{Kremer2023} suggested that a repeated TDE by a stellar-mass BH in a dense star cluster could explain the multi-wavelength emission of AT2018cow \cite[]{Kremer2019}. Although star-forming complexes were indeed found near the location of AT2018cow and are associated with the same host galaxy (as evidenced by the same redshift), the significantly higher extinction measured for these star-forming complexes compared to that of AT2018cow indicates that they are background objects located behind AT2018cow \citep[]{Sun2023}. Thus, there is currently insufficient evidence to support the presence of a dense star cluster environment for AT2018cow.}

One possible explanation is that AT2018cow is a failed massive star explosion event \citep{Margutti2019, Perley2019} in which the stellar core collapses, forming a compact object such as a magnetar or a stellar-mass {BH} \citep[]{Kashiyama2015, Margutti2019}. {Due to the huge energy release from a central source, in the form of either a magnetar wind or an accretion-disk wind, part of the in-falling envelop might be turned back, resulting in a fast-moving outflow. The supernova (SN) shock in massive stars like Wolf-Rayet stars (WRs) or blue supergiants (BSGs) will stall due to their tightly bound envelops and the steep density gradients therein, leading to the less mass loss driven by the SN shock (typically $\sim 10^{-2} - 10^{-4} \, M_{\odot}$) \citep[]{Kashiyama2015, Fernandez2018}. Any electromagnetic signals from a central engine would remain observable without significant obscuration \citep[]{Kashiyama2015}.}

{Note that red supergiants (RSGs) or yellow supergiants (YSGs) may also experience failed supernovae. However, due to their massive stellar envelopes, the supernova shock typically drives the ejecta with the masses of a few $M_{\odot}$ \citep[]{Kashiyama2015, Fernandez2018}. In such cases, the electromagnetic radiation from the central engine would be heavily obscured by the optically thick ejecta. As the shocked ejecta expand outward, they would produce the observational signatures of a typical supernova explosion, such as a lower peak luminosity ($\sim 10^{42} \, \rm{erg/s}$) and a slower  ($\sim \rm months$) evolution in light curves. At later times, narrow emission lines may emerge due to the interaction between the outflow and the dense ejecta. However, none of these predicted observational signatures were detected in AT2018cow \citep[]{Margutti2019, Perley2019}. Therefore, the progenitor of AT2018cow is more likely to be either a WR or a BSG.}

The following {evidences support} this hypothesis: (1) The properties of {its} host galaxy, such as the relationship between its star formation rate and galaxy mass, are similar to those of massive{-}star {explosive} events like LGRBs and core-collapse supernovae \citep[]{Ho2023}. (2) A quasi-periodic oscillation (QPO) signal at a frequency of 224Hz has been detected in the soft X-ray band of AT2018cow \citep[]{Pasham2022}. {Note however that a much slower QPO is also detected \citep[]{Zhang2022}, suggesting a much heavier {BH} might be plausible as well.} (3) Recent observations indicate that AT2018cow continues to emit UV and X-ray radiation even after 1400 days, suggesting the presence of a persistent radiation source at the center of AT2018cow \cite[]{Sun2022, Sun2023, Chenyy2023, Migliori2024}.

The failed massive star explosion scenario also provides a natural explanation to the multi-wavelength radiation of AT2018cow. The radio emission arises from the interaction between the high-density CSM, formed by stellar winds prior to the massive star explosion, and the fast-moving {outflow}. The bright X-ray radiation originates from the spin-down of a central magnetar or from the accretion onto a stellar-mass BH \citep[]{Margutti2019, Perley2019, Liu2022}.

Regarding the origin of the UV-optical-NIR radiation of AT2018cow, there are currently two possible models. One is the interaction between the {outflow} and the CSM \citep[]{Fox2019, Xiang2021}, while the other is the radiation from the central engine and then reprocessed by the outflow \citep[]{Piro2020, Uno2020, Uno2020b}. The former model fails to explain the broad line features observed from the early to late phases (full-width $\Delta \lambda \sim 1500 \, \rm{\AA} - 200 \, \AA$) in the spectrum of AT2018cow {\citep[]{Perley2019}}. On the contrary, the latter model can explain the broad line features as well as the rapid evolution of the photospheric radius of AT2018cow{, in which the} fast expanding outflow causes the photosphere to recede rapidly and produces a wide Doppler broadening in the emission line \citep[]{Perley2019}. The reprocessing model can also account for the NIR excess observed \citep[]{Margutti2019, Lu2020}. However, whether the central engine is a magnetar or a stellar-mass BH is still unknown.

In this paper, we aim to constrain the mass of the central compact object  {$M_{\rm obj}$} in order to speculate on the nature of {the} central engine. {By obtaining the mass of the outflow $M_{\rm out}$, we can estimate the mass of the central compact object via $M_{\rm obj} \approx M_{\rm pre} - M_{\rm out}$. Here, some empirical knowledge about the pre-explosion stellar mass $M_{\rm pre}$ has to be utilized. Studying the core-collapse of different types of massive stars,} \cite{Woosley2012} concludes that {they} have different pre-explosion masses: {$M_{\rm pre} \gtrsim 8 \, M_{\odot}$} for Wolf-Rayet stars, while {$M_{\rm pre} \gtrsim 20 \, M_{\odot}$} for blue supergiants. Since WRs eject the hydrogen-rich shells almost entirely during the late stages of their evolution, but the spectrum of AT2018cow exhibits strong hydrogen emission features, it is unlikely that the progenitor of AT2018cow was a WR and rather suggests that it was most likely a BSG. Thus we adopt $M_{\rm pre} \gtrsim 20 \, M_{\odot}$ for AT2018cow in this paper.

{Note that  we could rule out the scenario where AT2018cow's progenitor experienced significant mass loss prior to explosion. Significant mass loss would deplete the hydrogen envelope. In that case, one should expect to see no \textemdash \, or extremely weak  \textemdash \, broad-emission-lines of hydrogen during the explosion. However, strong broad-lines of hydrogen were detected throughout the entire evolution of AT2018cow \citep[]{Margutti2019, Perley2019}, indicating that its progenitor probably did not undergo significant mass loss.}

We will consider the reprocessing model of the outflow to estimate $M_{\rm out}$. The mass loss rate $\dot{M}$, the velocity $v_{\rm out}$, and the {internal energy density (represented by the temperature $T$) of the outflow} may shape the observed SED \citep[]{Margutti2019, Lu2020,Roth2016, Roth2020}. {\cite{Lu2020} {found} that the outflow reprocessing may result in a significant NIR excess in the observed SED.} By calculating the emitted spectrum based on the reprocessing model, we could fit the model results to the observed SED of AT2018cow. Subsequently we may obtain the best-fit parameters ($\dot{M}, \, v_{\rm out}, \, T$), and estimate $M_{\rm out}$.

Previous analytical work did not consider frequency-dependent opacity \citep[]{Loeb1997, Strubbe2009, Kashiyama2015, Piro2020}, while the computational cost associated with numerical calculations was prohibitively large \citep[]{Roth2016, Dai2018, Thomsen2022, Parkinson2022}, making direct application to observational data challenging. In this work, we perform analytical calculations of the reprocessing model considering the frequency-dependent opacities, {which enables a rapid computation of the emitted spectrum.}

In section \ref{Model}, we describe the reprocessing model. In section \ref{Application}, we apply our model to AT2018cow. In section \ref{Discussion}, we discuss the limitations of our model, and we summarize our results in section \ref{Conclusions}.

\section{Model} \label{Model}

\begin{figure}
\centering
\includegraphics[scale = 0.4]{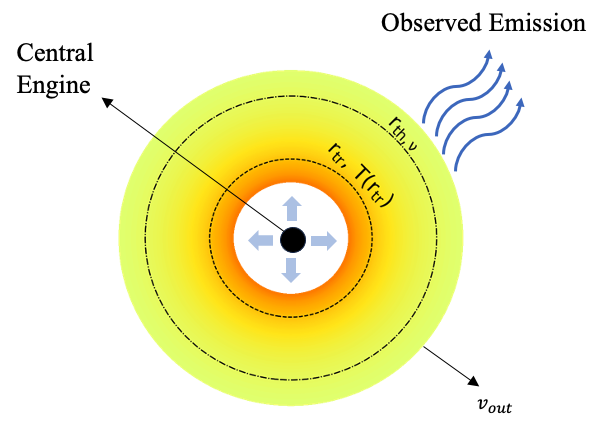}
\caption{Schematic of the reprocessing model. The photon trapping radius $r_{\rm tr}$ (Eq. \ref{trap_cal}) separates the outflow into two radial regions: the inner adiabatic-cooling dominated region ($r < r_{\rm tr}$), and the outer radiative-transport dominated region ($r > r_{\rm tr}$). The frequency-dependent thermalization radius $r_{\rm th, \nu}$ defines the last absorption radius for photons of frequency $\nu$ (Eq. \ref{eff_dep}). Only those photons emitted at $r > r_{\rm th, \nu}$ are not absorbed on its way out.}
\label{schematic_outflow}
\end{figure}

The radiative reprocessing model, in which an optically thick outflow absorbs the high-energy emission from a central source and re-emits in longer wavelengths, has been invoked to explain FBOTs \citep[]{Piro2020, Uno2020, Uno2020b}.  Here in this paper, we assume that the outflow is spherically symmetric and composed of pure hydrogen gas for simplicity. The outflow might be the accretion disk winds \citep[]{Kashiyama2015,Piro2020} or the magnetar winds \cite[]{Margutti2019}. Given the mass loss rate $\dot{M}$ and the outflow velocity $v_{\rm out}$, the density profile of the outflow could be roughly written as
\beq \label{density}
\rho (r, t) = \frac{\dot{M}(t)}{4 \pi r^2 v_{\rm out}},
\eeq
{The outflow region that interests us is $r \lesssim v_{\rm out} t$, thus we neglect the material travel time here.}

The radiation reprocessing could be treated separately in two radial regions of the outflow{, as} shown in Figure \ref{schematic_outflow}:\delete{adiabatic cooling in the inner region of the outflow, and radiative transport in the outer region of the outflow.}  {the inner adiabatic-cooling dominated region, and the outer radiative-transport dominated region.}

\subsection{Adiabatic Cooling Region}

Photons are injected from the inner boundary of the outflow, where the gas density is so high, photons are  "frozen" within the shell due to the electron scattering. {Here we consider the electron scattering opacity only, as it dominates the total opacity in highly ionized gas  \citep[]{Piro2020, Piro2025}.} The electron scattering optical depth from outside of the outflow  {to a radius $r$ inside} is
\beq
\begin{split}
\tau_{\rm es}(r)& = \int_{r}^{R_{\rm out}} \kappa_{\rm es} \rho(r) dr \\
&= \kappa_{\rm es} \frac{\dot{M}}{4 \pi v_{\rm out}} \left(\frac{1}{r} - \frac{1}{R_{\rm out}}  \right), 
\end{split}
\eeq
where $\kappa_{\rm es} = 0.4 \, \rm{cm^2 \, g^{-1}} $ is the electron scattering opacity for pure hydrogen gas, {$R_{\rm out} = R_{\rm in} + v_{\rm out} t$ and $R_{\rm in}$ are the outer and inner boundaries of the outflow, respectively.} For $r \ll R_{\rm out}$, $\tau_{\rm es}(r)$ could be roughly written as $\tau_{\rm es}(r) \approx \kappa_{\rm es} \rho(r) r$.
We adopt the formula in \cite{Piro2020} to estimate the photon diffusion time\delete{at a specific radius r}  {from $r$}:
\beq
t_{\rm dif} \simeq \frac{\tau_{\rm es}(r)}{c} \frac{(R_{\rm out} - r) r}{R_{\rm out}},
\eeq
which matches the expected limits: $t_{\rm dif} \approx \tau_{\rm es}(r) (R_{\rm out} - r) /c$ when $r \approx R_{\rm out}$, and $t_{\rm dif} \approx \tau_{\rm es}(r) r/c$ when  $r \ll R_{\rm out}$.

Further outward, we define the trapping radius $r_{\rm tr}$, below which ($r < r_{\rm tr}$) photons are trapped in the {moving} shell, while beyond $r_{\rm tr}$, the photons may escape from the local fluid \citep[]{Strubbe2009, Kashiyama2015, Piro2020, Chen2022}. {It is} determined by {equating the outflow dynamic time there to the photon diffusion time $t_{\rm dif}(r_{\rm tr})$} \citep[]{Piro2020}, or
\beq \label{trap_cal}
\frac{r_{\rm tr}- R_{\rm in}}{v_{\rm out}} \simeq \frac{\tau_{\rm es}(r_{\rm tr}) (R_{\rm out} -r_{\rm tr}) r_{\rm tr}}{c R_{\rm out}}.
\eeq
For $R_{\rm in} \ll r_{\rm tr} \ll R_{\rm out}$, Eq. \eqref{trap_cal} roughly gives $\tau_{\rm es}(r_{\rm tr}) \approx c/v_{\rm out}$, or
\beq \label{dif_rad}
r_{\rm tr} \approx \frac{\kappa_{\rm es} \dot{M} }{4\pi c} \approx 2.2 \times 10^{14} \, \rm{cm} \left(\frac{\dot{M}}{10^{-7} M_{\odot} \, s^{-1}} \right).
\eeq
 {Note that} Eq. \eqref{dif_rad} corresponds to the case of late times in \cite{Piro2020}. 

Below $r_{\rm tr}$, the radiation pressure $P = aT^4/3$ dominates the total pressure \citep[]{Strubbe2009}, where $T$ is the temperature, $a$ is the Boltzmann energy density constant. {As the shell moves outward from $R_{\rm in}$, the photons are adiabatically cooled until reaching $r_{\rm tr}$.}
The radiation energy density\delete{ in the shell } follows the adiabatic law as
\beq \label{ED}
aT(r)^4 \propto \rho(r)^{4/3}.
\eeq

As the shell reaches {$r_{\rm tr}$}, photons within the shell may {start to} diffuse out of the local fluid, with a diffusive luminosity given by:
\beq \label{Ldif}
L_{\rm dif} \simeq 4 \pi r_{\rm tr}^2 aT(r_{\rm tr})^4 v_{\rm out},
\eeq
Combining Eqs. \eqref{dif_rad} and \eqref{Ldif}, one obtains
\beq 
\begin{split} \label{Ldif_scale}
L_{\rm dif}& \simeq 1.4 \times 10^{41} \, \rm{erg \, s^{-1}} \left(\frac{\dot{M}}{10^{-7} M_{\odot} \, s^{-1}} \right)^2 \\
& \times \left(\frac{v_{\rm out}}{0.1 \, c} \right) \left[\frac{T(r_{\rm tr})}{10^{4} \, \rm{K}} \right]^{4} {.}
\end{split}
\eeq

\subsection{Radiative Transport Region}

Beyond $r_{\rm tr}$, photons diffuse out by radiative transport, {while they} might be scattered or absorbed {and re-emitted} in this region. The photons of different wavelengths could be last absorbed at different radii, which we define as the frequency-dependent thermalization radius\delete{(also the effective mean path)} $r_{\rm th, \nu}$ \citep[]{Rybicki1979}, at which
\beq \label{eff_dep}
 \sqrt{\tau_{\rm abs, \nu} (\tau_{\rm abs, \nu} + \tau_{\rm es})} =1,
\eeq
where 
\beq \label{abs_tot}
\tau_{\rm abs, \nu}  =  \int_{r_{\rm th, \nu}}^{R_{\rm out}} (\kappa_{\rm ff, \nu} + \kappa_{\rm bf, \nu} + \kappa_{\rm bb, \nu} ) \rho dr 
\eeq
is the frequency-dependent pure absorption optical depth,\delete{and} $\kappa_{\rm ff, \nu}$ is the free-free opacity, $\kappa_{\rm bf, \nu}$ is the bound-free opacity,  and $ \kappa_{\rm bb, \nu}$ is the bound-bound opacity {for pure hydrogen gas}. We {will} neglect the bound-bound opacity hereafter since we aim to fit the model to the observed SED, so the line features are not considered. {It is also reasonable to neglect the bound-bound opacity of hydrogen and the absorption opacities of metals, as they have a very weak impact on the total opacity in the optical-NIR bands for highly ionized gas \citep[]{Roth2016}.}

Since the electron scattering opacity ($\kappa_{\rm es}$) dominates the total opacity, the observed spectrum could be roughly given by the chromatic radiative diffusion equation \citep[]{Illarionov1972,Rutten2003,Shen2015,Lu2020} as
\beq \label{spec_de}
\begin{aligned}
L_{\nu} &\approx -4\pi r^2 \frac{4 \pi \partial B_{\nu} [T(r)]}{\partial [\kappa_{\rm es} + \kappa_{\rm ff, \nu}(r) + \kappa_{\rm bf, \nu}(r)] r} \\
& \approx  -4\pi r^2 \frac{4 \pi \partial B_{\nu} [T(r)]}{\partial \tau_{\rm es}(r)}. 
\end{aligned}
\eeq
Note that Eq. \eqref{spec_de} gives the observed spectrum when it is applied to the characteristic radius $r_c = \rm{max}$$(r_{\rm tr}, r_{\rm th, \nu})$, because it is from this radius onward that the energy of a photon would not be changed any more. Therefore, using an approximation to the derivative term in Eq. \eqref{spec_de}, we rewrite {it} as
\beq  \label{spec_def}
L_{\nu} \simeq 4\pi r_c^2 \times \frac{4 \pi B_{\nu}[T(r_c)]}{\tau_{\rm es}(r_c)}.
\eeq

The opacities are determined by the gas density and the temperature {(see below)}. The temperature profile for $r > r_{\rm tr}$ is obtained by solving the {bolometric} radiative transport equation:
\beq \label{RTE}
\frac{d aT^4}{dr} = -\frac{3 \kappa_{\rm es} \rho(r)}{4 \pi c r^2} L_{\rm dif},
\eeq
which gives
\beq \label{T_rdt}
T(r) \simeq \left[\frac{\kappa_{\rm es} \dot{M} L_{\rm dif}}{4 \pi r^3 ac \times 4 \pi v_{\rm out}} \right]^{1/4}, \,  \mbox{for} ~ r  > r_{\rm tr},
\eeq
where $L_{\rm dif}$ is given by Eq. \eqref{Ldif} and is $\simeq \int L_{\nu} d\nu$.

\subsection{Absorptive Opacities}

Here we describe the frequency-dependent opacities that we used. The following are constants used: $e$, $m_e$, $m_p$, $h$, $k_B$, $\sigma_T$ are the unit charge,  the mass of the electron, the mass of the proton, the Planck constant, the Boltzmann constant, the electron scattering cross-section{,} respectively.

\subsubsection{{F}ree-free opacity}
For free-free transition, the \delete{ free-free absorption} opacity \delete{{$\kappa_{\rm ff, \nu}$}} ($\rm cm^{-1}$) is \citep[]{Rybicki1979}
\beq \label{ff_abs}
\begin{aligned}
\kappa_{\rm ff, \nu}& = \frac{4 e^6}{3 m_e h c} \left(\frac{2 \pi}{3k_B m_e} \right)^{1/2} T^{-1/2} Z^2 n_e n_i \nu^{-3} \\
&\times (1-e^{-h\nu/kT}) \rho^{-1} g_{\rm ff},
\end{aligned}
\eeq
where \delete{$\alpha_{\rm ff, \nu}$ is the free-free absorption coefficient,} $Z$ is the net charge, $n_e$, $n_i$ are the number density of the electrons and the ions, and $g_{\rm ff}$ is the Gaunt factor. In the Rayleigh-Jeans limit ($h\nu \ll k_B T$), and neglect the Gaunt factor, Eq. \eqref{ff_abs} becomes
\beq \label{alpha_ff}
\kappa_{\rm ff, \nu} = 0.018 T^{-3/2}   Z^2 \nu^{-2} n_e n_i \rho^{-1}. 
\eeq

\subsubsection{{B}ound-free opacity}
For pure hydrogen gas, the bound-free opacity is approximately given by \citep[]{Osterbrock2006}
\beq \label{kappa_bf}
\kappa_{\rm bf, \nu} \approx \frac{\sigma_{\rm bf, s}(\nu) f_n}{Z m_p},
\eeq
where $f_n$ is the neutral fraction. For pure hydrogen gas, we have $Z = 1$. The photoionization cross-section $\sigma_{\rm bf, s} (\nu)$ could be given as \citep[]{Rybicki1979,Roth2016}:
\beq \label{sigma_bf_s}
\sigma_{\rm bf, s} (\nu) = N_s \sigma_0 \left(\frac{h\nu}{\chi_i} \right)^{-3},
\eeq
where $N_s$ is the principal quantum number in the energy level $s$, $\chi_i$ is the ionization potential, $\sigma_0 = 6.3 \times 10^{-18} \rm{cm}^2$ for H \citep[]{Roth2016}.

Next, we consider the neutral fraction $f_n$. The photoionization equilibrium gives \citep[]{Osterbrock2006,Metzger2014,Metzger2016}
\beq \label{pe}
n_{\rm HI, s} \int_{\nu_{\rm thr}}^{\infty} \frac{4 \pi J_{\nu}}{h \nu} \sigma_{\rm bf, s}(\nu) d\nu = n_{\rm HII} n_e \alpha_{\rm rec, s}(T),
\eeq
where $J_{\nu}$ is the mean intensity of radiation ($\rm{erg \ s^{-1} \ cm^{-2} \ ster^{-1} \nu^{-1}}$), ${4 \pi J_{\nu}} / {h \nu}$ is the number of the incident photons ($\rm{s^{-1} \ cm^{-2} \nu^{-1}}$), $\nu_{\rm thr}$ is the photoionization threshold frequency, and $n_{\rm HI, s}$ is the number density of neutral hydrogen in different states, $n_{\rm HII}$ is the number density of $\rm{HII}$, and $\alpha_{\rm rec, s}(T)$ is the recombination coefficient \footnote{The recombination coefficient data could be downloaded from \url{https://www.astronomy.ohio-state.edu/nahar.1/nahar_radiativeatomicdata/}.}. It then gives
\beq \label{f_n}
\begin{split}
f_n &= \frac{n_{\rm HI, s}}{n_{\rm HI, s} + n_{\rm HII}} \\
&= \left(1 + \frac{4 \pi}{\alpha_{\rm rec} n_e}  \int_{\nu_{\rm thr}}^{\infty} \frac{J_{\nu}}{h\nu} \sigma_{\rm bf, s} d\nu  \right)^{-1}.
\end{split}
\eeq
The bound-free opacity could be obtained by substituting Eqs. \eqref{sigma_bf_s} and \eqref{f_n} into Eq. \eqref{kappa_bf}.
 
\begin{figure}
\centering
\includegraphics[scale = 0.3]{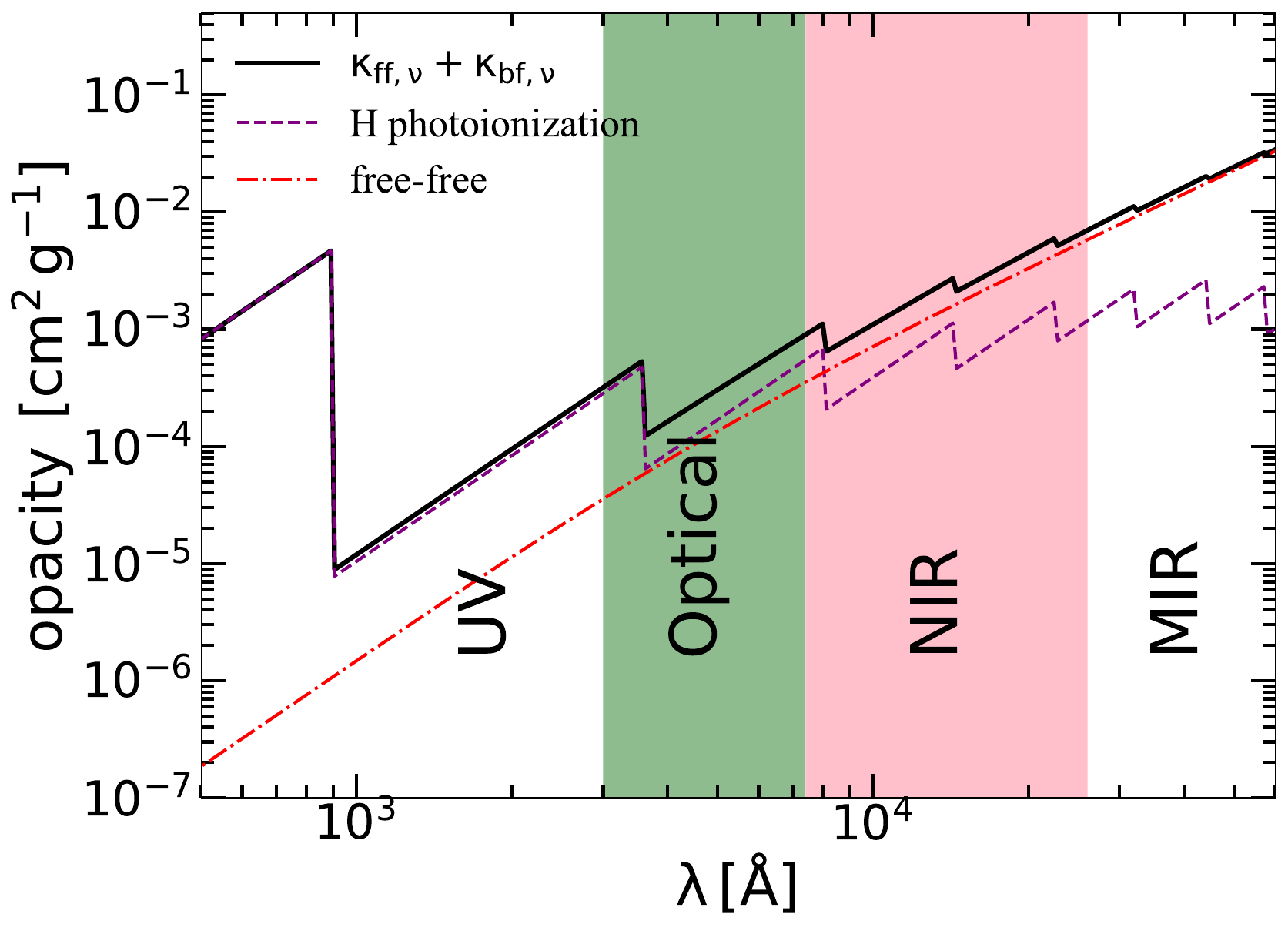}
\caption{The opacities for the bound-free photoionization, and the free-free transitions at the different wavelengths for pure hydrogen gas at $\rho = 7.6 \times 10^{-14} \, \rm{cm \, g^{-1}}$, $T = 2.2 \times 10^4 \, \rm{K}$ (note these conditions correspond to the density and temperature at $r_{\rm tr}$ for the numerical example in Figure \ref{fig:specpng}). In the NIR band and even longer wavelengths, the absorptive opacity is dominated by $\kappa_{\rm ff, \nu}$. We neglect the bound-bound opacity here since we aim to fit the model results to the observed SED, so the line features are ignored.}
\label{kappa_1127}
\end{figure}

Figure \ref{kappa_1127} shows the free-free opacity $\kappa_{\rm ff, \nu}$ and bound-free opacity $\kappa_{\rm bf, \nu}$ for a given density and temperature as an example. At low frequencies, $\kappa_{\rm ff, \nu}$ dominates the absorptive opacity, while at high frequencies, $\kappa_{\rm ff, \nu}$ dominates. This fact could be utilized to simplify Eq. \eqref{eff_dep} in order to get an asymptotic expression for $r_{\rm th, \nu}$  {[see Eq. \eqref{rthnu}]}.

\subsection{Spectral Shape of NIR Band} \label{2.4}

In this part, we will briefly derive the analytical form of the spectrum under the Rayleigh-Jeans limit ($h\nu \ll k_B T$) \citep[]{Chandrasekhar1950,Shakura1969,Felten1972,Roth2020}.  As shown in Figure \ref{kappa_1127}, in the NIR band, the total absorptive opacity $\kappa_{\rm abs, \nu}$ is\delete{higher and} dominated by the $\kappa_{\rm ff, \nu}$, i.e., $\kappa_{\rm abs, \nu} \approx \kappa_{\rm ff, \nu}$. Note that $\kappa_{\rm es} \gg \kappa_{\rm ff, \nu}$, Eq. \eqref{eff_dep} becomes
\beq \label{sou_rthnu}
\sqrt{\tau_{\rm ff, \nu} \tau_{\rm es}} \approx 1,
\eeq
 where $\tau_{\rm ff, \nu} = \int_{r}^{R_{\rm out}} \kappa_{\rm ff, \nu} \rho(r) dr \approx \kappa_{\rm ff, \nu} \rho(r) r$, and $\tau_{\rm es} \approx \kappa_{\rm es} \rho(r) r$. Solving Eq. \eqref{sou_rthnu} using {Eqs. \eqref{density}, \eqref{T_rdt} and \eqref{alpha_ff}}, one could {get} a solution for $r_{\rm th, \nu}$ \delete{that satisfies the}  {in the} Rayleigh-Jeans {limit}
\beq \label{rthnu}
\begin{split}
& r_{\rm th, \nu }  \simeq 3.5 \times 10^{14} \, {\rm{cm}} \, \left(\frac{v_{\rm out}}{0.1 \, c } \right)^{-3/4} \left[\frac{T(r_{\rm th,\nu})}{10^4 \, \rm{K}} \right]^{-3/8} \\
&\times \left( \frac{\dot{M}}{10^{-7} \rm{M_{\odot} \, s^{-1}}} \right)^{3/4}   \left(\frac{\nu}{3 \times 10^{14} \, \rm{Hz}} \right)^{-1/2},
\end{split}
\eeq
which matches the results in \cite{Lu2020} and \cite{Roth2020}.

In the NIR band, $r_{\rm th, \nu} \propto \nu^{-1/2}$ as in Eq. \eqref{rthnu}. \delete{For}  {At} lower frequencies \delete{, which satisfy}  {where} $r_{\rm th, \nu} > r_{\rm tr}$, the monochromatic luminosity is given by Eq. \eqref{spec_def} {with $r_c = r_{\rm th, \nu}$}. {Then} using Eq. \eqref{rthnu}, we could obtain the analytical form of the spectrum in the NIR band:
\beq \label{spec_num}
\begin{split}
&\lambda L_{\lambda}  \simeq 2.4 \times 10^{40} \, {\rm{erg \, s^{-1}}} \left( \frac{\dot{M}}{10^{-7} \rm{M_{\odot} \, s^{-1}}} \right)^{5/4}  \\
&  \times \left(\frac{v_{\rm out}}{0.1 \, c} \right)^{-5/4} \left[\frac{T(r_{\rm th,\nu})}{10^4 \, \rm{K}} \right]^{-1/8} \left(\frac{\lambda}{10^{4} \, \rm{\AA}} \right)^{-3/2}.
\end{split}
\eeq
Since the observed NIR luminosity {depends} weakly on the temperature, Eq. \eqref{spec_num} could be \delete{applied to measure the relationship between} {used to infer} the outflow mass loss rate and \delete{the outflow} velocity ($\dot{M}, \, v_{\rm out}$)  {from the observed spectrum}.

\begin{figure} 
\centering
\includegraphics[scale = 0.35]{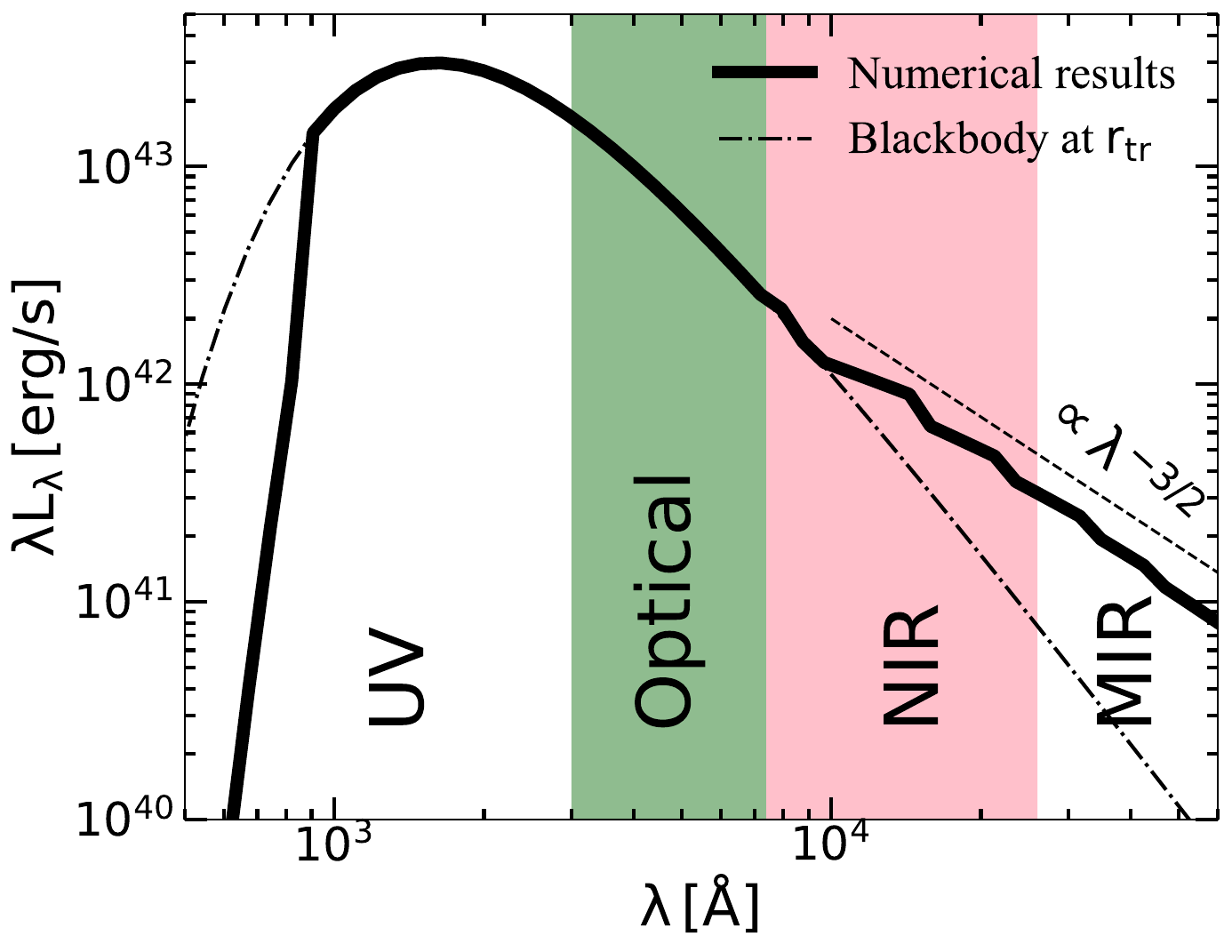}
\caption{An example of the emergent spectrum from a reprocessing outflow, numerically calculated from Eq. \eqref{spec_def}. The parameters are set as $\dot{M} = 10^{-6} \, \rm{M_{\odot} \, s^{-1}} $, $v_{\rm out} = 2 \times 10^{9} \, \rm{cm \, s^{-1}}$, $T(r_{\rm tr})= 2 \times 10^{4} \, \rm{K}$. The black dash-dotted line is the black-body spectrum whose temperature is $T(r_{\rm tr})$. The dashed line is the asymptotic shape (Eq. \ref{spec_num}) for the NIR excess. The UV sharp dropoff at $\lambda \lesssim 913 \AA$ is due to the absorptive opacity being dominated by hydrogen $\kappa_{\rm bf, \nu}$ there (see Figure \ref{kappa_1127}), which leads to an increase of $r_{\rm th, \nu}$, hence a lower temperature. Although the radiating area increases, the lower $T$ results in a decrease in the radiative intensity $B_{\nu} [T(r_{\rm th, \nu})]$ at these wavelengths, ultimately causing a significant drop in $L_{\nu}$ (Eq. \ref{spec_def}).}
\label{fig:specpng}
\end{figure}

Figure \ref{fig:specpng} is a numerical example of the emitted spectrum. {The major shape of the spectrum is that of a blackbody with the temperature set at $T(r_{\rm tr})$, except that at the lower and higher frequency ends, there are a NIR excess and a UV drop-off, respectively. In the intermediate wavelength range of $\sim 1000 \, \AA - 7000 \, \AA$,} since the absorptive opacity is low there (see Figure \ref{kappa_1127}) such that $r_{\rm th, \nu} < r_{\rm tr}$, photons in this wavelength range could escape from $r_{\rm tr}$ without being absorbed; so their spectrum is in a blackbody shape given by Eq. \eqref{spec_def} with $r_c = r_{\rm tr}$. Therefore, $T(r_{\rm tr})$ would correspond to the color temperature of the observed optical/UV SED.

The spectrum in the NIR band follows the asymptotic form described in Eq. \eqref{spec_num}. The results indicate that photons emitted at $r_{\rm tr}$ and of the frequencies that satisfy $r_{\rm th, \nu} > r_{\rm tr}$ {would} be absorbed on their way out. The frequency dependence (Figure \ref{kappa_1127} and Eq. \ref{rthnu}) of $r_{\rm th, \nu}$ {suggests that the lower-frequency photons would have a larger emission area ($> 4\pi r_{\rm tr}^2$), which, according to Eq. \eqref{spec_def}, results in a higher {luminosity $L_{\nu} > L_{\nu}[T(r_{\rm tr})]$}. Therefore, the spectrum exhibits a significant NIR excess whose shape is $\lambda L_{\lambda} \propto \lambda^{-3/2}$ (Eq. \ref{spec_num})}, deviating from the Rayleigh-Jeans shape.


\section{Application to AT2018cow} \label{Application}
\begin{figure*}[h]
\centering
\includegraphics[scale = 0.25]{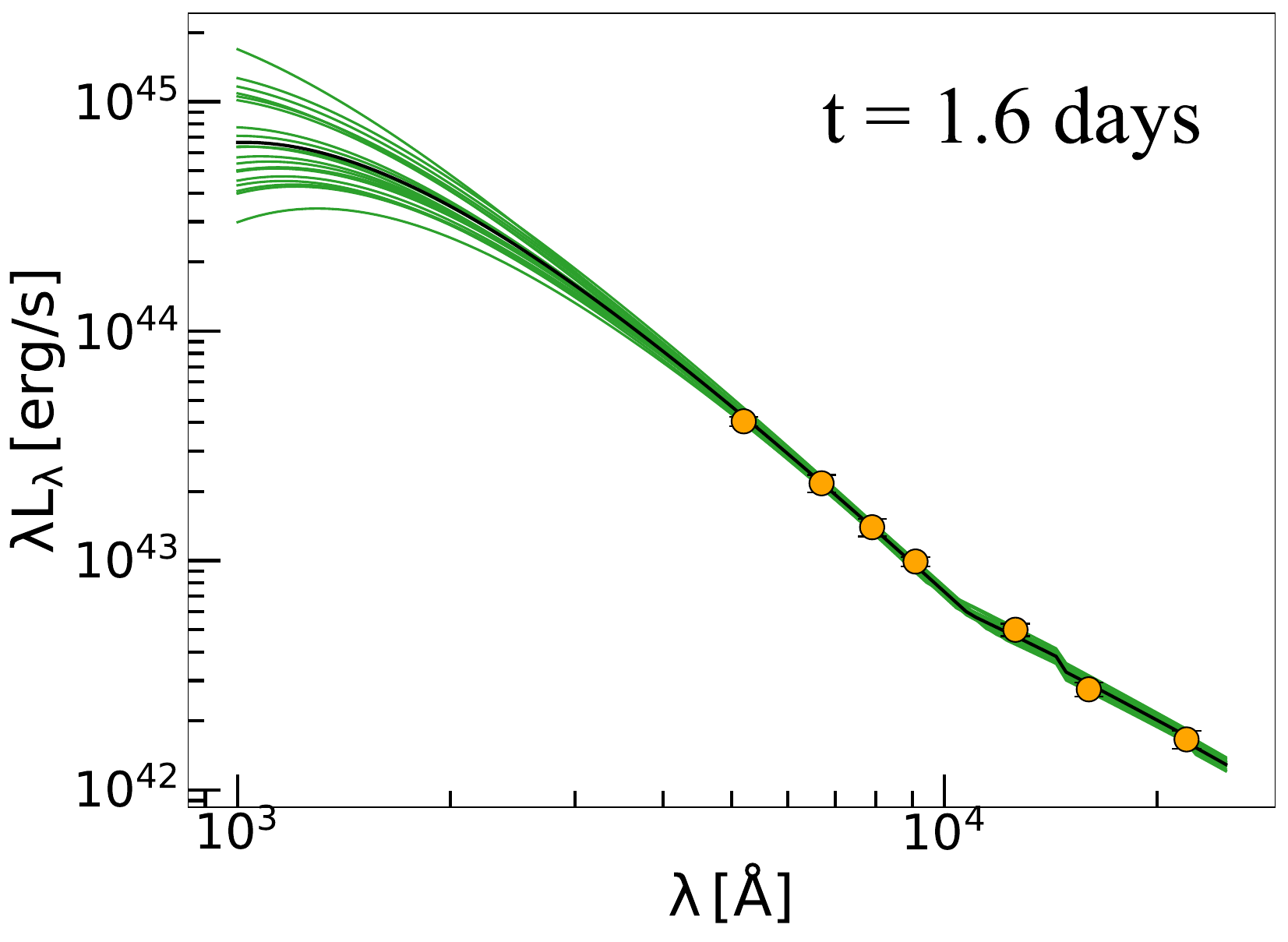}
\includegraphics[scale = 0.25]{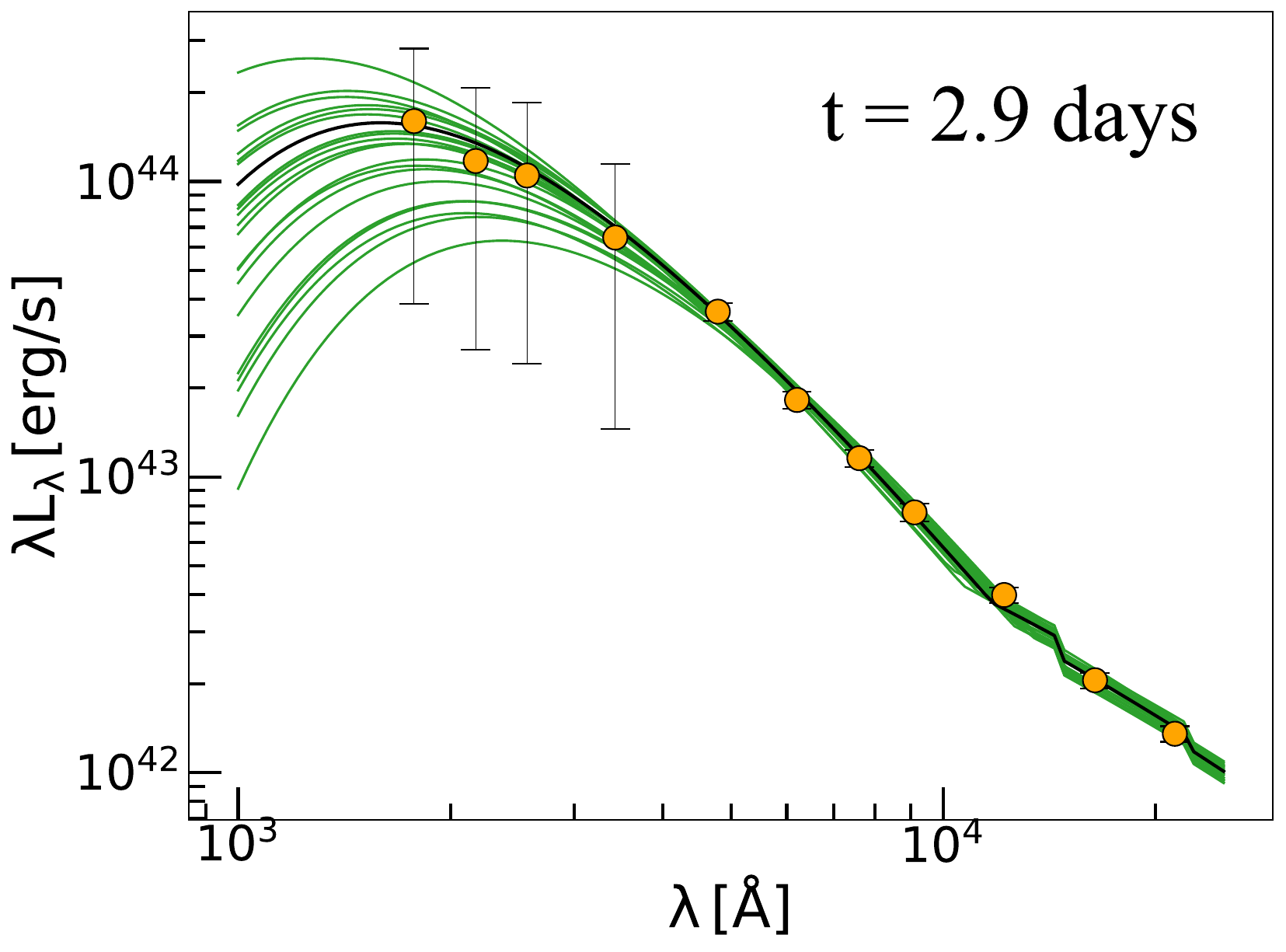}
\includegraphics[scale = 0.25]{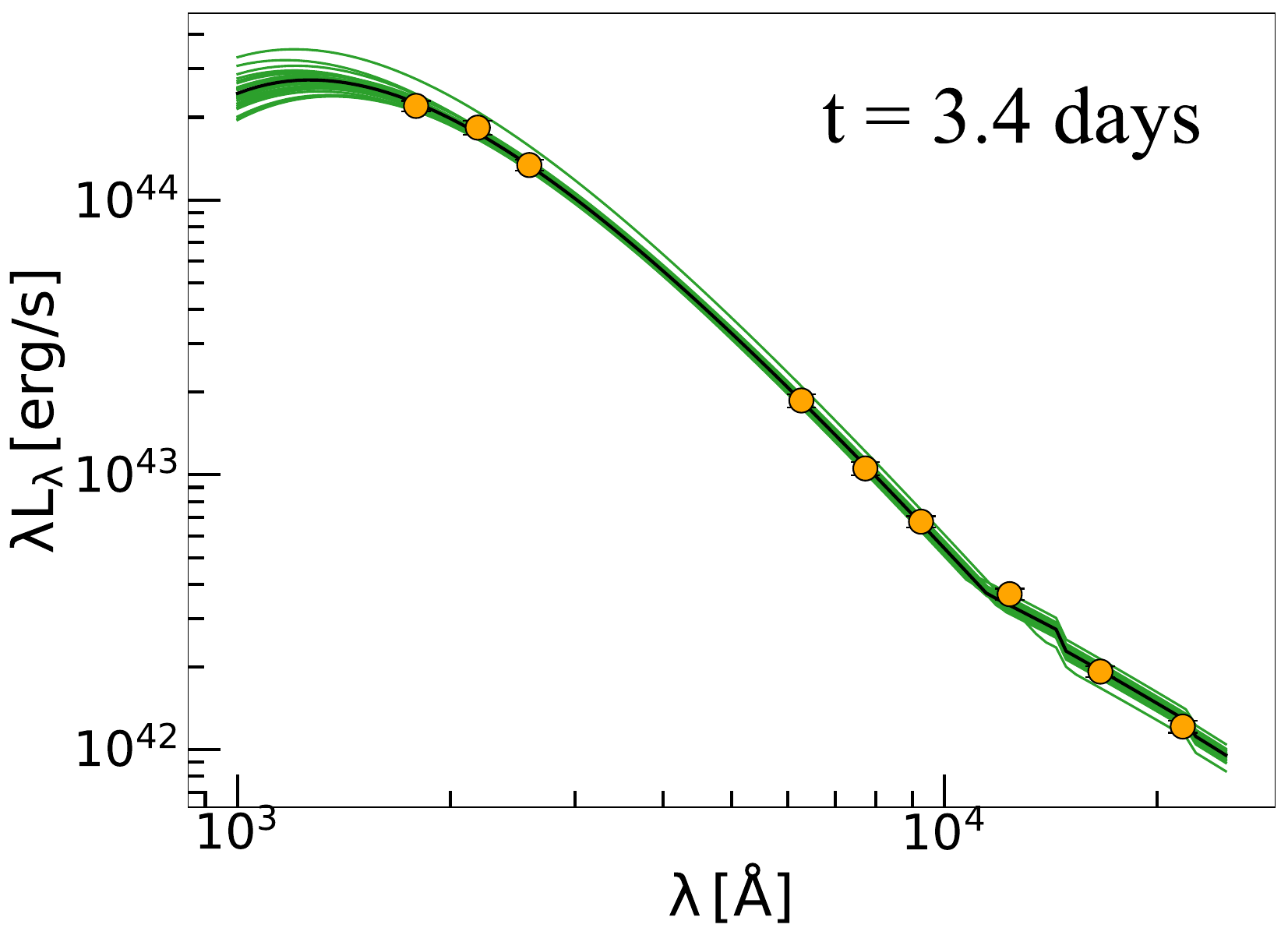}
\includegraphics[scale = 0.25]{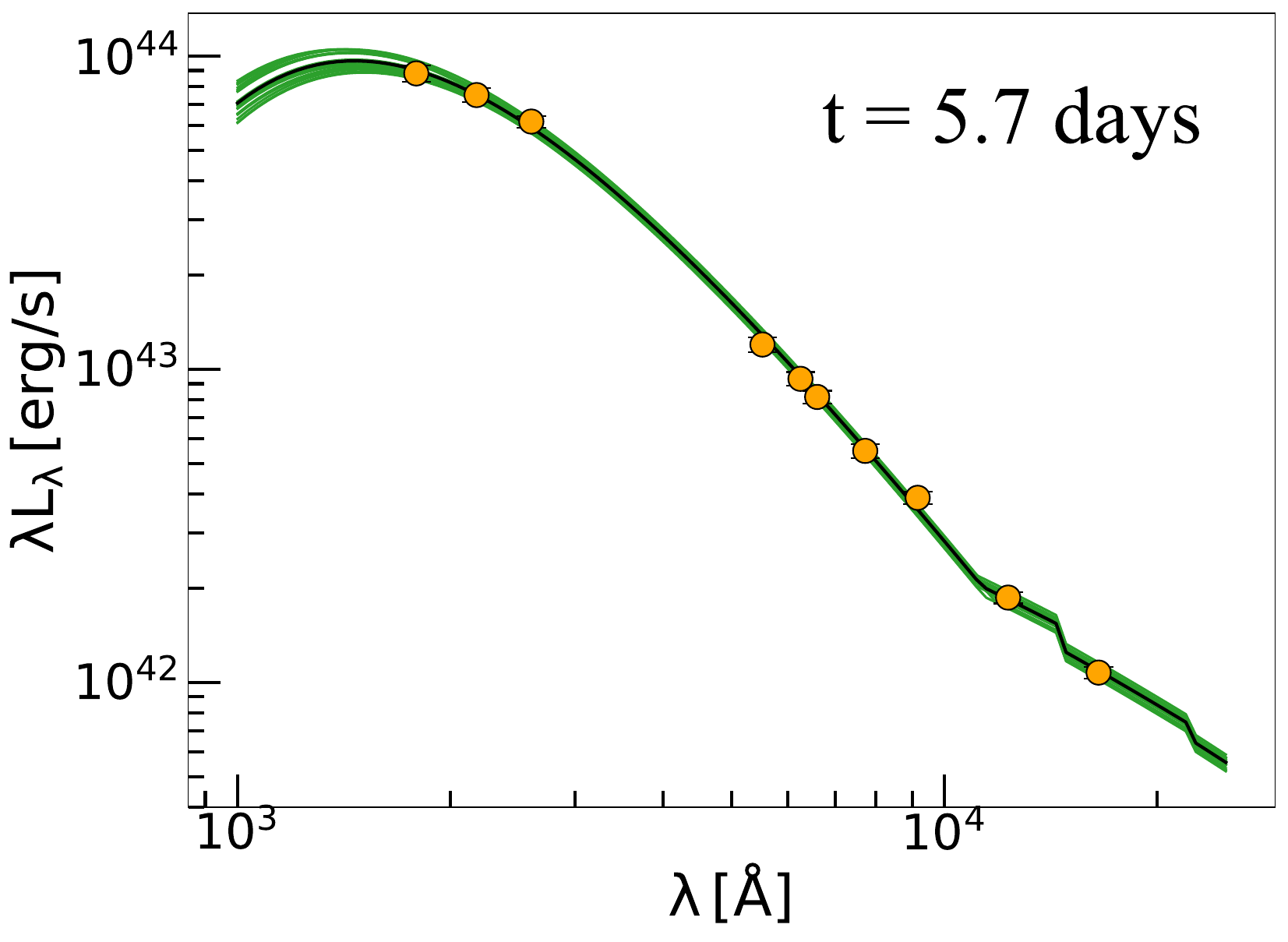} 
\includegraphics[scale = 0.25]{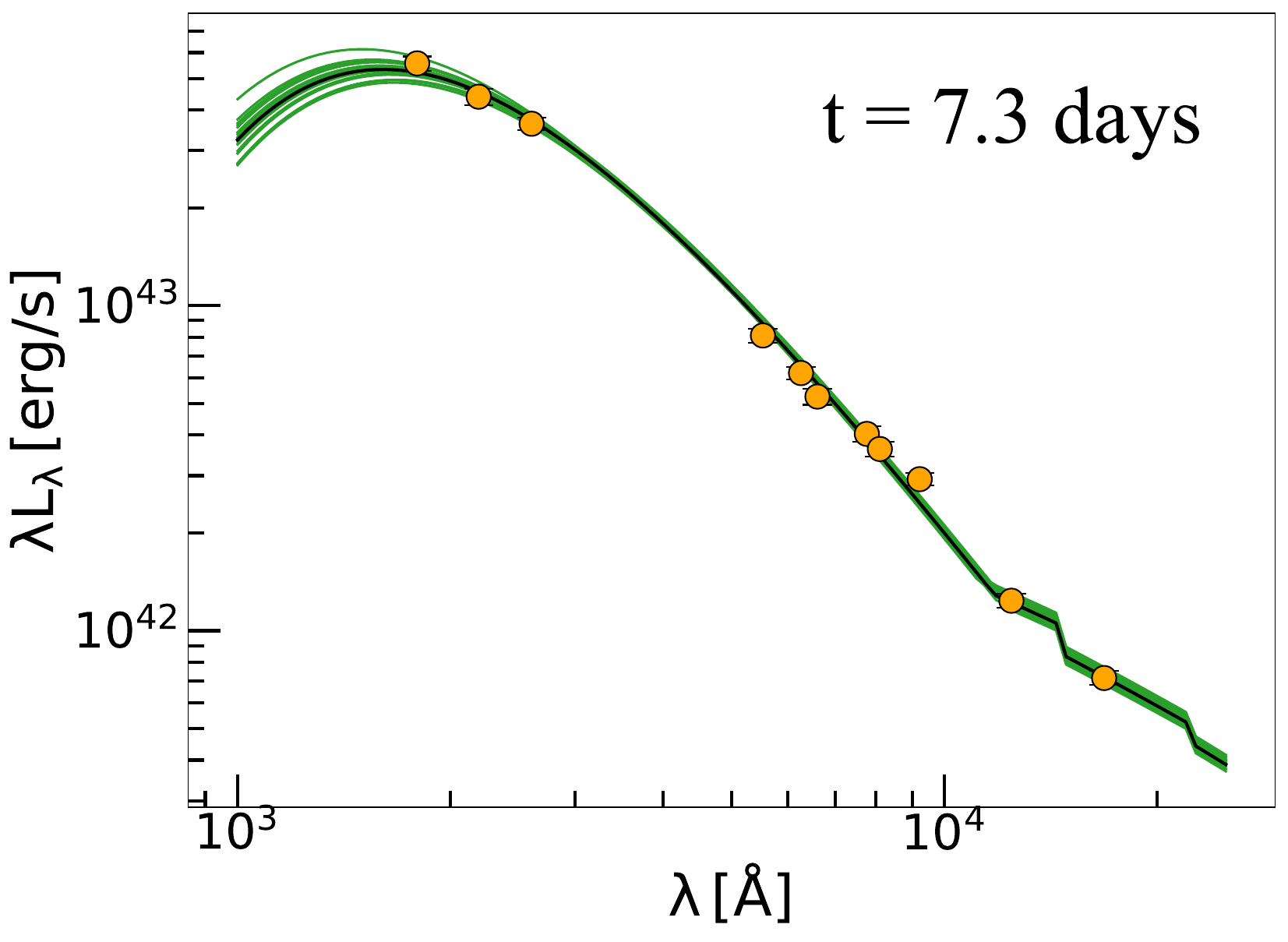}    
\includegraphics[scale = 0.25]{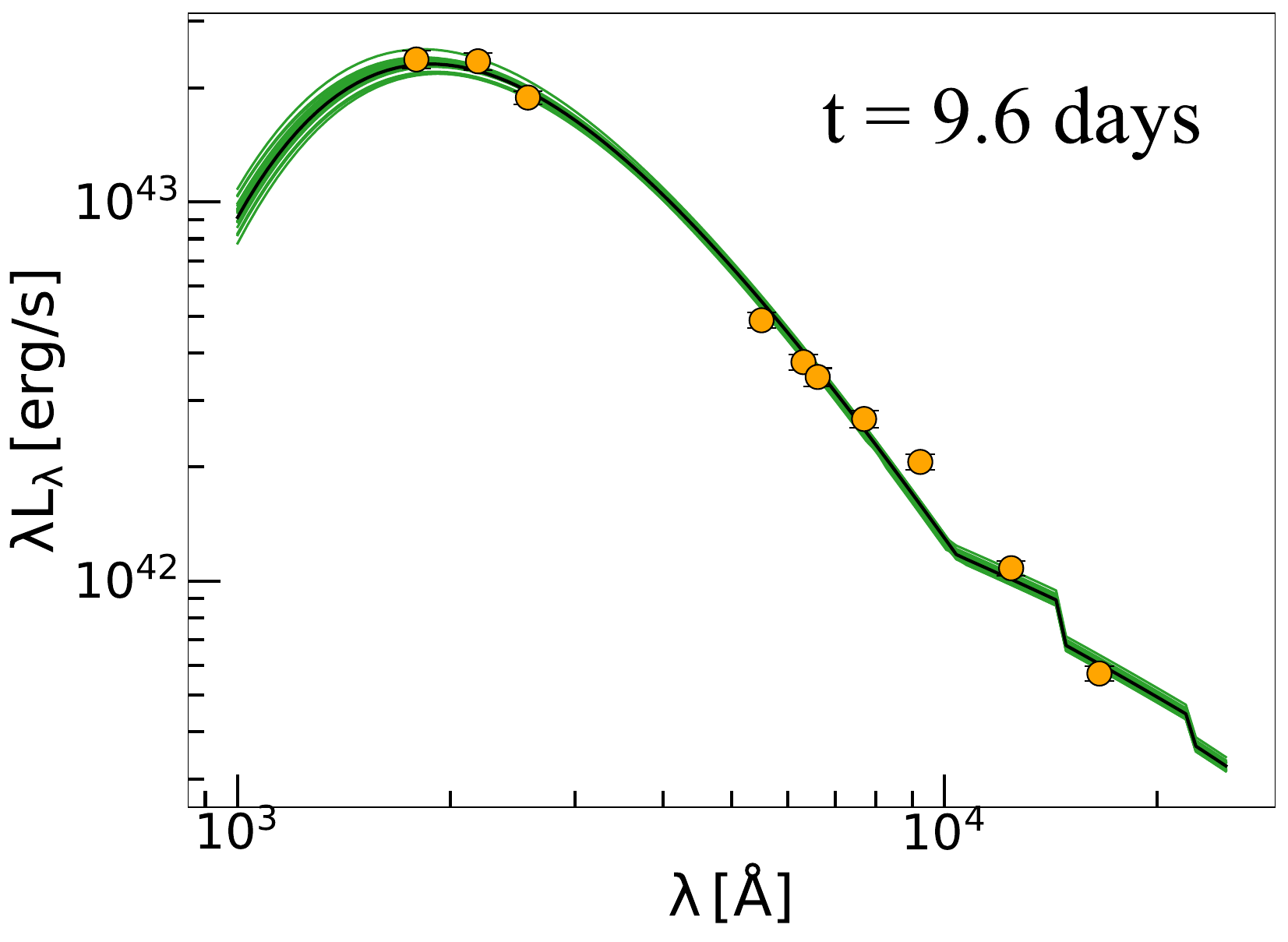} 
\includegraphics[scale = 0.25]{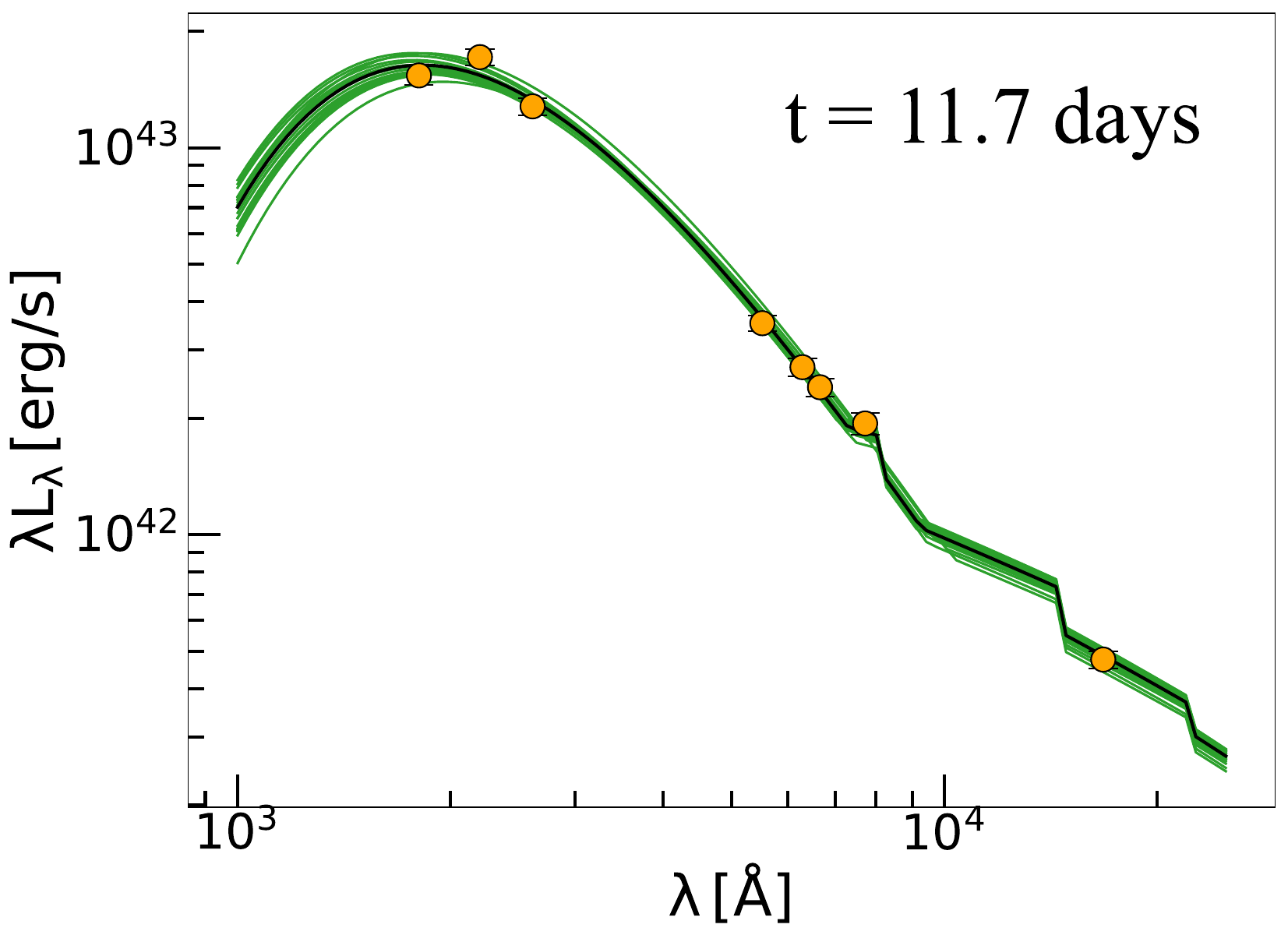} 
\includegraphics[scale = 0.25]{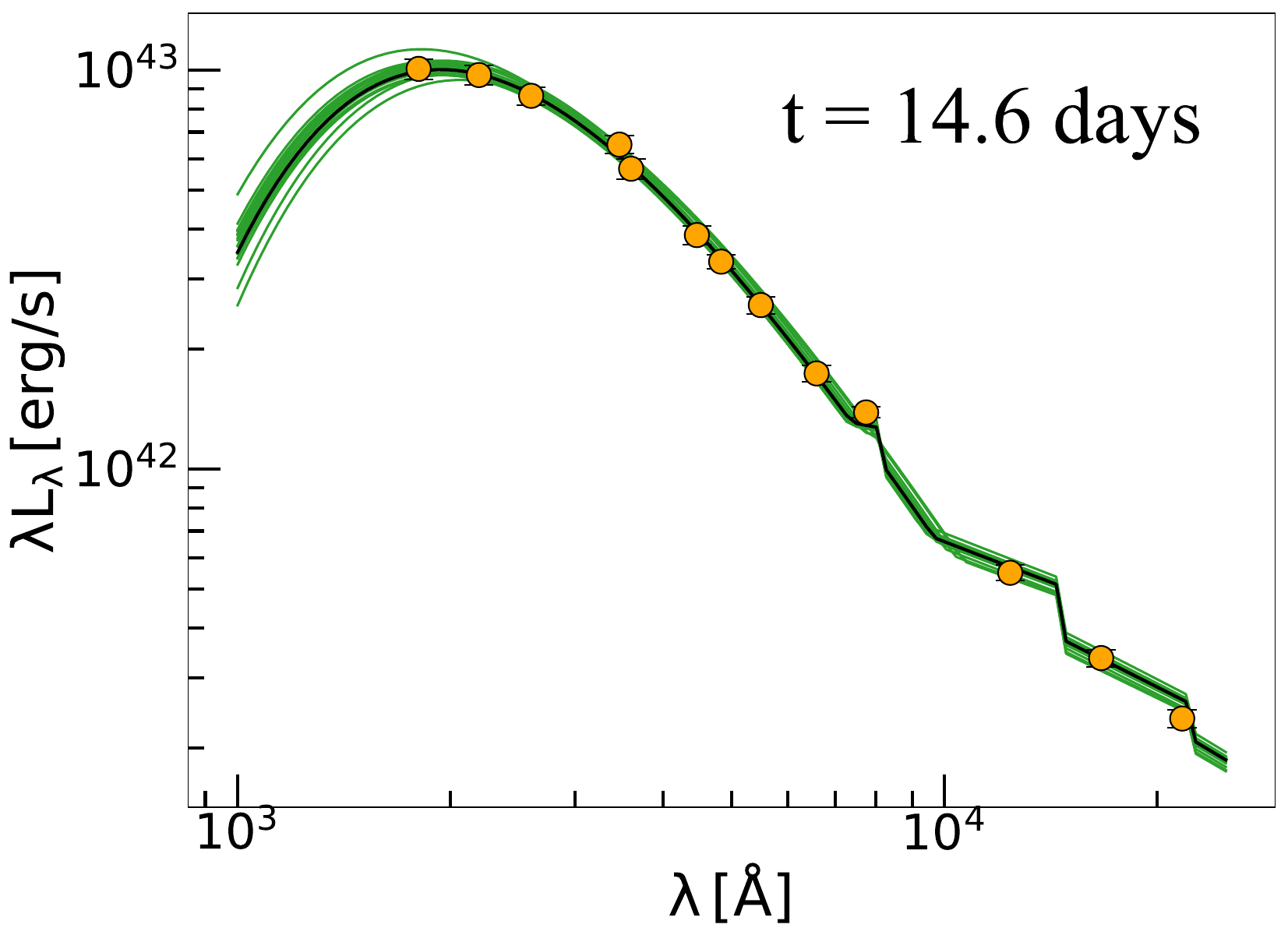} 
\caption{The results of fitting the reprocessing outflow model to the SED data of AT2018cow at eight observing epochs ($t = $ 1.6 to 14.6 days) \citep[]{Perley2019}. The black solid lines are the best-fit results, and the green solid lines correspond to the 1$\sigma$ uncertainty. We set the time $t = 0$ as the first detection of AT2018cow, $MJD58285$ in the ATLAS $o$-band \citep[]{Perley2019}. {Note that there were no UV band observations for $t < 3.0$ days. The UV band data at $t = 2.9$ days were obtained through a temporal extrapolation from the $t = 3.0$ day data, which might introduce a systematic error. Thus for the $t = 2.9$ day SED UV-band data, we introduce a systematic error equivalent to $10 \%$ of the observed value.}}
\label{cowfit_rin1e9}
\end{figure*}

\begin{figure}[h]
\centering
\includegraphics[scale = 0.4]{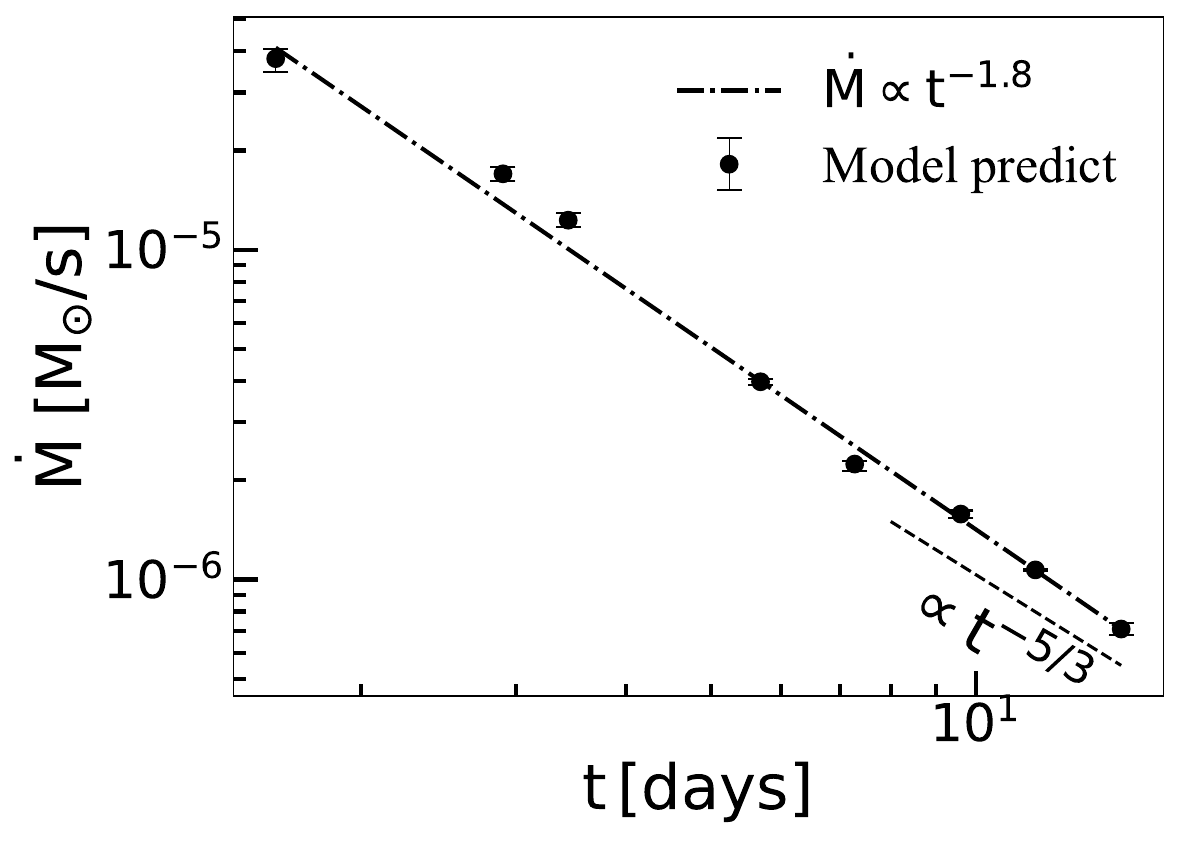}
\includegraphics[scale = 0.4]{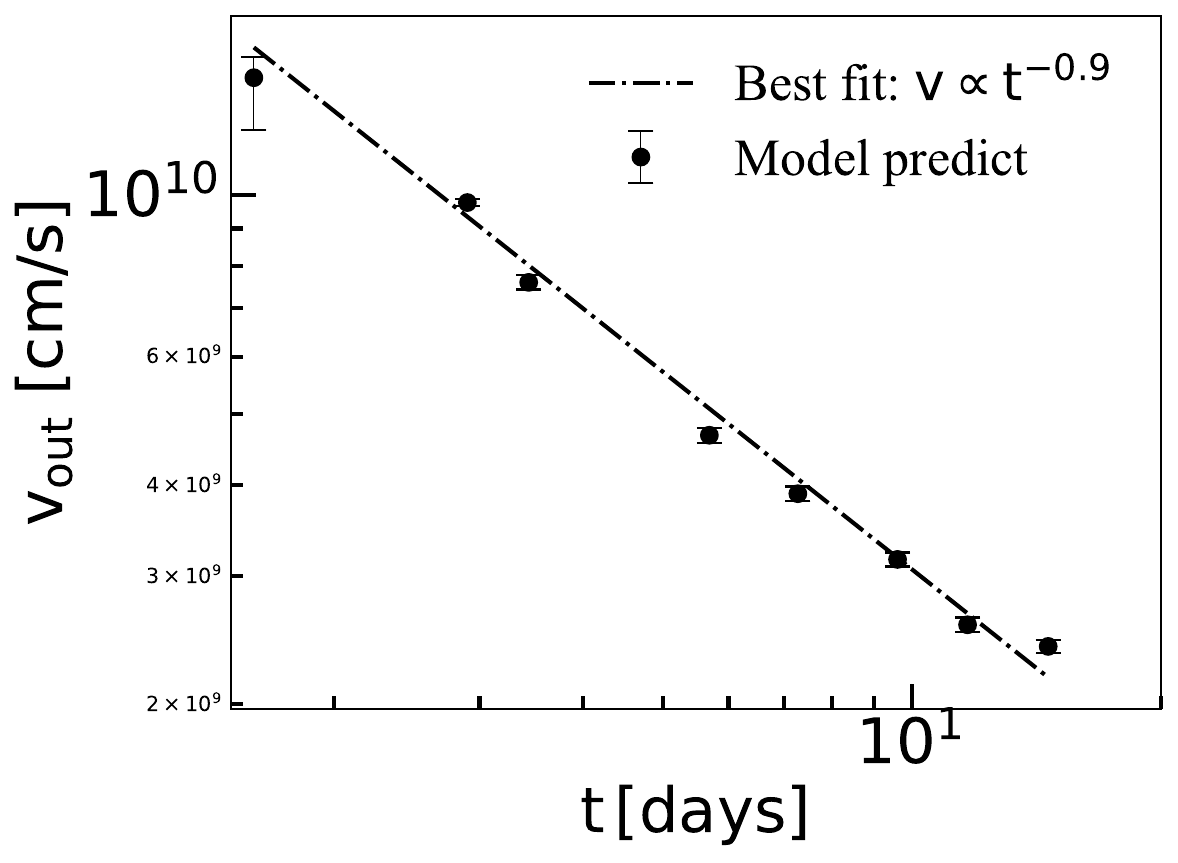}
\includegraphics[scale = 0.4]{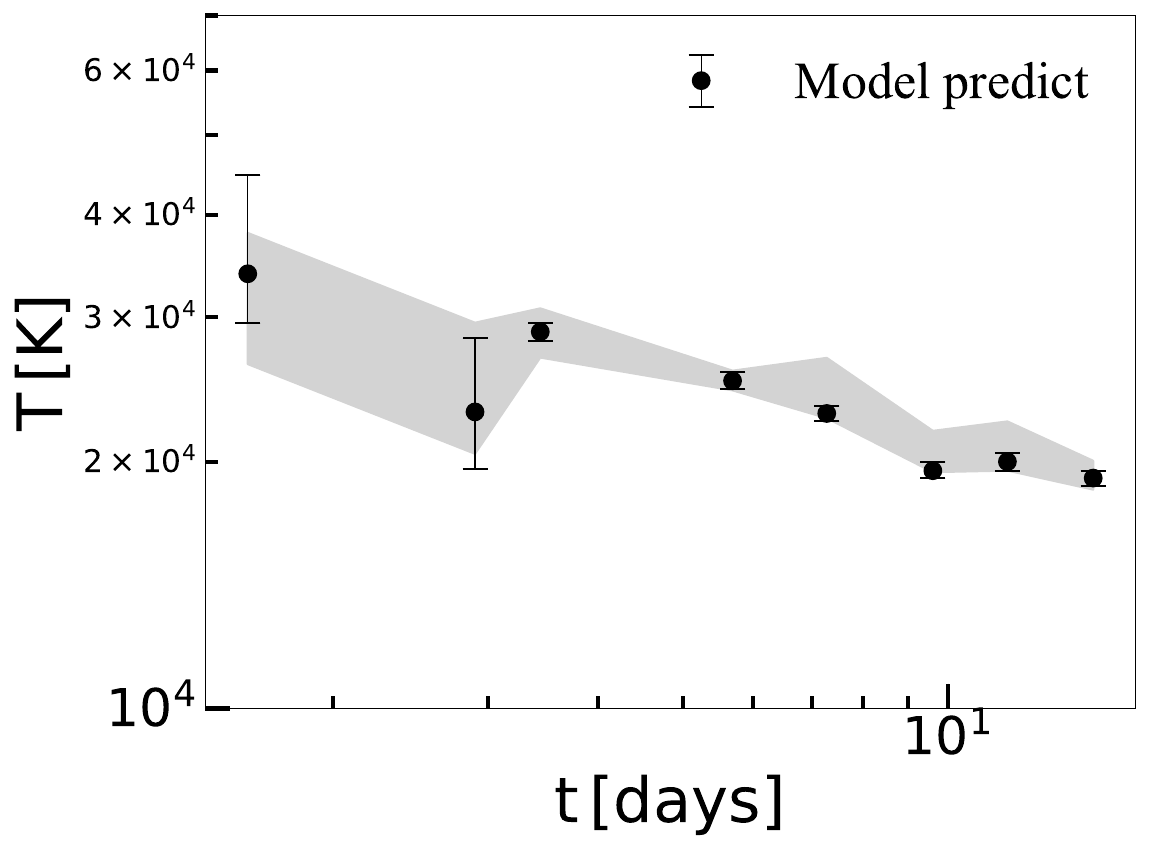}
\caption{The best-fit parameters obtained from fitting the outflow reprocessing model to the SED data of AT2018cow. The top panel shows the evolution of the mass loss rate $\dot{M}$. The middle panel shows the evolution of the outflow velocity $v_{\rm out}$. The bottom panel shows the evolution of $T(r_{\rm tr})$. The data points are best-fit parameters. The dash-dotted lines are the {power-law function fit to the data points.} The grey region marks the temperature obtained by fitting the SED to blackbody (BB) $+$ power law model with 1 $\sigma$ confidence. The first data point has exceptionally large error bars due to  the lack of observations in the UV band at $t = 1.6$ days.}
\label{cowfit_para}
\end{figure}

Using the numerical model we described in {S}ection \ref{Model}, we can obtain the outflow parameters ($\dot{M}$, $v_{\rm out}$, $T$) by fitting it to the SED data of AT2018cow. {It} has  {UV/optical/NIR} SED data over about a few $\times 10 \, \rm{days}$ \citep[]{Prentice2018, Perley2019}, {from which we use} the SED data taken {from $t = $} 1.6 to 14.6 days, {as shown in Figure \ref{cowfit_rin1e9}.}

\subsection{Utilizing the NIR break}

{It} exhibits a break in the NIR bands with a significant NIR excess \citep[]{Perley2019}. The break frequency $\nu_{\rm b}$ is different at different days. \cite{Metzger2023} used the dust echo model to interpret such NIR excess. {Here} we apply the outflow reprocessing model to explain it\delete{\citep[]{Margutti2019}} {\citep[]{Roth2016, Metzger2016, Dai2018, Lu2020, Piro2020}}. 

In our work, $\nu_{\rm b}$ in the NIR bands is roughly given by $r_{\rm th, \nu} = r_{\rm tr}$  {(see Section \ref{2.4})}, which {in turn contains model parameters $\dot{M}$, $v_{\rm out}$, etc}. Combining Eqs. \eqref{dif_rad} and \eqref{rthnu}, we could obtain the break wavelength:
\beq \label{lambda_b}
\begin{split}
\lambda_{\rm b}&  \equiv \frac{c}{\nu_{\rm b}} \simeq 10^{3}  \,\rm{\AA} \left(\frac{\dot{M}}{10^{-7} \, \rm{M_{\odot} \, s^{-1}}}  \right)^{1/2} \\
& \times \left(\frac{v_{\rm out}}{0.1 \, c} \right)^{3/2} \left[\frac{T(r_{\rm tr})}{10^{4} \, K} \right]^{3/4}.
\end{split}
\eeq
{From Eq. \eqref{spec_num}, the monochromatic luminosity at $\lambda_{\rm b}$ is given by}:
\beq \label{spec_num1}
\begin{split}
&\lambda_{\rm b} L_{\lambda_{\rm b}}  \simeq 2.4 \times 10^{40} \, {\rm{erg \, s^{-1}}} \left( \frac{\dot{M}}{10^{-7} \rm{M_{\odot} \, s^{-1}}} \right)^{5/4}  \\
& \times \left(\frac{v_{\rm out}}{0.1 \, c} \right)^{-5/4} \left[\frac{T(r_{\rm tr})}{10^4 \, \rm{K}} \right]^{-1/8} \left(\frac{\lambda_{\rm b}}{10^{4} \, \rm{\AA}} \right)^{-3/2}.
\end{split}
\eeq
Here, $T(r_{\rm tr})$ is the color temperature. Since [$\lambda_{\rm b}$, $\lambda_{\rm b} L_{\lambda_{\rm b}}$, $T(r_{\rm tr})$] could be obtained from the observation, the outflow parameters ($\dot{M}, \, v_{\rm out}$) could be {quickly} estimated by combining Eqs. \eqref{lambda_b} and \eqref{spec_num1} as
\beq \label{v_out}
\begin{split}
v_{\rm out}& \simeq 6.6 \times 10^9 \, \rm{cm \, s^{-1}} \left(\frac{\lambda_{\rm b}}{10^4 \, \AA} \right)^{1/5} \\
& \times \left(\frac{\lambda_{\rm b} L_{\rm b} }{10^{42} \, erg \, s^{-1}}\right)^{-1/5} \left[\frac{T(r_{\rm tr})}{10^{4} \, K} \right]^{-2/5},
\end{split}
\eeq
and
\beq \label{Mdot}
\begin{split}
\dot{M}& \simeq 1.0 \times 10^{-6} \, \rm{M_{\odot} \, s^{-1}} \left(\frac{\lambda_{\rm b}}{10^4 \, \AA} \right)^{6/5} \\
& \times \left(\frac{\lambda_{\rm b} L_{\lambda_{\rm b}}}{10^{42} \, \rm{erg \, s^{-1}}} \right)^{4/5} \left[\frac{T(r_{\rm tr})}{10^{4} \, K} \right]^{1/10}.
\end{split}
\eeq

For the early times of the SED ($t \lesssim 10 \, \rm{days}$), the observations give $\lambda_{\rm b} \simeq 15000 \, \AA$, $\lambda L_{\lambda} (\lambda \approx 15000 \, \AA) \simeq  2 \times 10^{42} \, \rm{erg \, s^{-1}}$ and $T_{\rm tr} \simeq 2 \times 10^4 \rm{K}$ \citep[]{Perley2019}. We can estimate the outflow parameters ($\dot{M}, \, v_{\rm out}$) using Eqs. \eqref{spec_num} and \eqref{lambda_b} at this stage as $v_{\rm out} \sim 5.0 \times 10^9 \, \rm{cm \, s^{-1}}$ and $\dot{M} \sim 1.9 \times 10^{-6} \, \rm{M_{\odot} \, s^{-1}}$. For the late times of AT2018cow, we have  $\lambda_{\rm b} \simeq 7000 \, \AA$, $\lambda L_{\lambda} (\lambda \approx 7000 \, \AA) \simeq 10^{42} \, \rm{erg \, s^{-1}}$ and $T_{\rm tr} \simeq 2 \times 10^4 \rm{K}$ \citep[]{Perley2019}{, which give} $v_{\rm out} \sim 4.5 \times 10^{9} \, \rm{cm \, s^{-1}}$ and $\dot{M} \sim  7.0 \times 10^{-7} \, \rm{M_{\odot} \, s^{-1}}$.

{Note that most other FBOTs, unlike AT2018cow, may have only the optical and near-UV SED data and lack the NIR data, thus without showing the break. For them, Eqs. \eqref{v_out} and \eqref{Mdot} could not be used to estimate the outflow parameters ($\dot{M}$, $v_{\rm out}$). In this case, two observables from the SED that one can utilize are the color temperature $T(r_{\rm tr})$ and the bolometric luminosity $L_{\rm dif}$. They provide a constraining relation between $\dot{M}$ and $v_{\rm out}$ via Eq. \eqref{Ldif_scale}. However, to determine $\dot{M}$, one has to obtain $v_{\rm out}$ independently, e.g., from measuring the width of the broad line features in the high-resolution spectra.}

\subsection{SED Fitting}

Alternatively in a holistic manner, we can fit the entire observed SED of AT2018cow to obtain more accurate outflow parameters, so that we could estimate the total mass of the outflow. Here we set [$\dot{M}$, $v_{\rm out}$, $T(r_{\rm tr})$] as the free parameters to fit the multi-epoch SED's of AT2018cow, using a MCMC package\footnote{Refer to \url{https://emcee.readthedocs.io/en/stable/} for details.}. The ranges of the free parameters are set as: $-9 < log_{10}(\dot{M}) < -1$; $3 < log_{10}[T(r_{\rm tr})] < 5$; $6< log_{10}(v_{\rm out}) < 10.2$. The fitting results are shown in Figure \ref{cowfit_rin1e9}. The best-fit parameters with $1\sigma$ confidence errors are listed in Table \ref{best-fit parameters_rin1e9}, and Figure \ref{cowfit_para} shows the evolution of [$\dot{M}, \, v_{\rm out}, \, T(r_{\rm tr})$].

The mass loss rate we obtained is nearly an order of magnitude larger than the results in \cite{Uno2020} and \cite{Piro2020} (both adopt the outflow reprocessing model) at $t \lesssim 5 \, \rm{days}$. In this paper, {we consider frequency-dependent absorption opacities, while \cite{Uno2020} and \cite{Piro2020} adopted constant (``gray'') opacity approximation.} 

\begin{table}[!ht] 
\centering 
\caption{Best-fit parameters with $1\sigma$ confidence errors from fitting AT2018cow's SED.}
\begin{tabular}{c|ccc} \hline 
$t \, (\rm{days)}$ & $\dot{M} \, \rm{(10^{-6} \, M_{\odot}/s)}$ & $v_{\rm out} \, \rm{(10^{9} \, cm/s)}$  & $T(r_{\rm tr}) \, \rm{(10^{4} \, K)}$   \\ \hline
1.6    & $38^{+3}_{-3} $ & $14^{+1}_{-2} $ & $3.4^{+1.1}_{-0.4} $ \\
2.9    & $16^{+2}_{-2} $ & $9.8^{+0.1}_{-0.1}  $ & $2.3^{+0.5}_{-0.3} $ \\
3.4    & $12^{+1}_{-1} $ & $7.6^{+0.2}_{-0.2}  $ & $2.9^{+0.1}_{-0.1} $   \\ 
5.7  & $4.0^{+0.1}_{-0.1}   $ & $4.7^{+0.1}_{-0.1}  $ & $2.5^{+0.1}_{-0.1}   $ \\ 
7.3  & $2.2^{+0.1}_{-0.1}   $ & $3.8^{+0.1}_{-0.1}  $ & $2.3^{+0.1}_{-0.1}   $ \\ 
9.6  & $1.6^{+0.1}_{-0.1}   $ & $3.2^{+0.1}_{-0.1}  $ & $2.0^{+0.1}_{-0.1}   $  \\ 
11.7 & $1.1^{+0.1}_{-0.1}  $ & $2.6^{+0.1}_{-0.1}  $ & $2.0^{+0.1}_{-0.1}   $ \\  
14.6 & $0.71^{+0.03}_{-0.03}  $ & $2.3^{+0.1}_{-0.1}  $ & $1.9^{+0.1}_{-0.1}   $ \\  \hline
\end{tabular}
\label{best-fit parameters_rin1e9}
\end{table}

By integrating the mass loss rate ($\dot{M}$), we could roughly estimate the total mass of the outflow $M_{\rm out}$:
\beq \label{M_out}
M_{\rm out} \simeq \int^{t_{\rm end}}_{t_{\rm start}} \dot{M} dt \approx 5.7^{+0.4}_{-0.4} \, M_{\odot}
\eeq
where $t_{\rm start} = 1.6 \, \rm{days}$ is the time of the first data point, and $t_{\rm end} = 14.6 \, \rm{days}$ is the time of the last data point. Note that the above estimate has neglected the early ($t < 1.6 \, \rm{days}$) and late ($t > 14.6  \, \rm{days}$) ejections of the outflow\delete {. Estimations of the ejection rates during the early and late phases can affect our estimation of the total mass of the outflow. Due}  {, due} to the lack of early SED data of AT2018cow there.

We obtained the outflow {velocity} $v_{\rm out} \approx 0.1 \, c \,- \,0.3 \, c$ {during} $t \, =  \, 1.6 - 14.6$ days as shown in the middle panel of Figure \ref{cowfit_para}. Note that \cite{Margutti2019} estimated lower {values} of the dense outflow's velocity, $\approx 4000$ km/s, based on the emission lines observed in the spectrum of AT2018cow {at $\gtrsim 20$ days}. Nevertheless, \delete {in} the early-time ($t \lesssim 20 \, \rm{days}$) spectrum of AT2018cow shows exceptionally broad emission lines (full-width $\Delta \lambda \sim 200 \, \rm{\AA} - 1500 \, \AA$), which corresponds to $v_{\rm out} \sim (1-7.5) \times 10^9 \, \rm{cm/s}$ \citep[]{Perley2019}. {Since our results suggest the SED generating $v_{\rm out}$ drops with time (Figure \ref{cowfit_para}), the high velocities we found for the early-time outflow are reasonable.}

\subsection{Implication on the Central Object}

Since we know the mass of the massive {BSG progenitor} before the explosion ($M_{\rm pre, SN} \gtrsim 20 \, M_{\odot}$ ), and we have obtained the total outflow mass $M_{\rm out}$, subtracting the two will allow us to estimate the mass of the central   {remnant} compact object $M_{\rm obj}$:
\beq \label{M_obj}
M_{\rm obj} = M_{\rm pre, SN} - M_{\rm out} \gtrsim 14  \, M_{\odot}.
\eeq
The upper limit of the mass of a neutron star is estimated to be $\sim \, 3.2 M_{\odot}$ \citep[]{Bombaci1996, Woosley2002}. Therefore, Eq. \eqref{M_obj} implies that the central compact object of AT2018cow is most likely a stellar-mass {BH}.

Interestingly, note that the evolution of the {inferred} mass loss rate $\dot{M}$ approximately follows a power-law behavior as $\dot{M} \propto t^{-5/3}$, which is commonly\delete { observed in}  {predicted for} the accretion processes of {the fallback material onto a} compact object {in a failed supernova scenario} \citep[]{Michel1988, Zhang2008, Dexter2013}. This {loosely} suggests the possibility of {fallback} accretion\delete {occurring} onto {a} stellar-mass {BH} in AT2018cow at early times.

\section{Discussion} \label{Discussion}

The outflow reprocessing model explains the observed SED of AT2018cow well. {Our model would be most applicable to those FBOTs that show (1) persistent and luminous X-ray emission, (2) NIR excess.} 

However, our model indeed has some limitations. Firstly, we assumed that the outflow is isotropic. {For} AT2018cow, as mentioned in Section \ref{Introduction}, there is evidence suggesting that the outflow is non-isotropic, with higher density concentrated in the equatorial region and lower density in the polar region \citep[]{Margutti2019}. Ignoring this angular dependence in our calculations could lead to discrepancies in the estimated mass of the outflow\delete {and the central engine}. If\delete {we also assume that} the outflow {was} concentrated within a specific solid angle $\Delta \Omega$, {the real outflow mass} $M_{\rm out}$ {would be lower} by a factor of $\Delta \Omega / 4 \pi$ {than our above estimate}. According to Eq. \eqref{M_obj}, a lower $M_{\rm out}$ would lead to a higher estimated mass of the central {object} $M_{\rm obj}$, which further supports the conclusion that the central engine of AT2018cow is most likely a stellar-mass {BH}.

{Secondly, given the model parameters $\dot{M}(t)$ and $v_{\rm out}$$(t)$ from the SED fitting, we estimate the total released energy $E_{\rm tot}$ during t = 1.6 to 14.6 days of the central engine. We could estimate $E_{\rm tot}$ through \citep[]{Shen2016}
\beq \label{Etot}
\begin{split}
E_{\rm tot}  \approx \int^{t_{\rm end}}_{t_{\rm start}}  (\frac{\dot{M} v^2}{2} + 4 \pi r_{\rm tr}^2 aT_{\rm tr}^4 v + \frac{G M \dot{M}}{R_{\rm in}} ) dt \\
 \approx \left[10^{53} + 2 \times 10^{52} \times \left(\frac{R_{\rm in}}{10^9 \, \rm{cm}} \right)^{-1} \right] \, \rm{erg}
\end{split}
\eeq
where $\int^{t_{\rm end}}_{t_{\rm start}}  ({\dot{M} v^2}/{2} + 4 \pi r_{\rm tr}^2 aT_{\rm tr}^4 v ) dt \approx 10^{53} \, \rm erg$ is estimated from our SED-fitting, and ${G M \dot{M}}/{R_{\rm in}}$ denotes the increasing in the gravitational potential energy of the shell as it moves from $R_{\rm in}$ to $r_{\rm tr}$ by assuming $R_{\rm in} \ll r_{\rm tr}$.} 

{The result in Eq. \eqref{Etot} implies that $E_{\rm tot} $ is larger than the total energy released in typical explosive events of a massive star (usually $\sim \, 10^{51} - 10^{52} \, \rm{erg}$) \citep[]{Smartt2009, Janka2012}. Considering the asymmetry of the outflow would reduce the energy estimate by only a factor of a few, not by many orders of magnitude. Therefore, this indicates that the central engine of AT2018cow might be an additional energy source. The total energy released via spin-down from a nascent magnetar is typically $\sim 10^{52}$ erg \citep[]{Thompson2004, Kaspi2017}, which is significantly lower than the minimum energy requirement from the central engine in our model. Taking into account our constraints on the mass of the central engine, we conclude that an accreting stellar-mass BH is the most likely energy source driving the outflow.}

Thirdly, we assume that the outflow is composed of pure hydrogen gas, which is not a rigorous treatment. The metallicity of the outflow is likely dependent on the mass and metallicity of the progenitor, making it difficult to simply estimate\delete {the metallicity of the outflow}. However, in the low-frequency band, the opacity is dominated by free-free opacity {for highly ionized outflow}, while the bound-free opacity of metal elements is relatively low \citep[]{Roth2016, Lu2020}. Therefore, the emitted {spectra} we obtained in the low-frequency band could still be considered reliable.

{Fourthly, in this paper we fit the observed early SED of AT2018cow from $t =$ 1.6 to 14.6 days only. This is because our results indicate that the majority of the mass loss occurred during the early stages. When we apply our model to fit the SED for $t >$ 20 days, we encounter difficulties since our model predicts a NIR SED shape of $\lambda L_{\lambda} \propto \lambda^{-1.5}$, which significantly deviates from AT2018cow's observed SED ($\lambda L_{\lambda} \propto \lambda^{0}$) during this phase. Remarkably, the observed NIR luminosity even exceeds the optical luminosity at $t > 40$ days. We infer that at $t >$ 20 days, as the temperature of the outflow decreases, the bound-bound and bound-free opacities of metals could no longer be neglected. In fact, when the gas temperature drops below 12,000 K, the outflow is no longer highly ionized, and absorptive opacities of metals may dominate the opacities even though their abundance is very low \citep[]{Uno2020}. At this stage, our model is no longer applicable. The emergence of the He I lines at $\sim$ 15 days also supports this inference \citep[]{Perley2019}.}

Finally, the emitted spectra we obtained are approximate results from analytical calculations. A more {rigorous} method is numerical simulation. Currently,  {sophisticated} codes such as Tardis \citep[]{Kerzendorf2014}, Sedona \citep[]{Kasen2006}, and PYTHON \citep[]{Long2002} are used for numerically computing the radiative transfer of the outflow utilizing Monte Carlo methods. However, this approach requires significant computational resources. In contrast, our model provides a rapid estimation of the emitted spectra, allowing us to fit our results to the observed SED and estimate the outflow parameters ($\dot{M}, \, v_{\rm out}$), in an adaptive and efficient way.

\section{Conclusions} \label{Conclusions}
In this paper, we invoke an outflow reprocessing model to explain the observed SED with significant NIR excess of AT2018cow. In our model, we consider the photon trapping ($r < r_{\rm tr}$) and the radiative diffusion ($r > r_{\rm tr}$) within the outflow. The observed photons originate from the radius {$\simeq$ max($r_{\rm tr}, \, r_{\rm th, \nu}$),} where the photons were last absorbed. In determining this radius, we consider the frequency-dependent opacities.

We calculate the SED from the outflow, and the results indicate that the emitted spectrum deviates from the blackbody shape both in the NIR band and the UV band. The spectrum exhibits a significant NIR excess, and follows the shape $\lambda L_{\lambda} \propto \lambda^{-3/2}$ in the NIR band. At the start {of} this excess, the break wavelength $\lambda_{\rm b}$ corresponds to $r_{\rm tr} = r_{\rm th, \nu_{\rm b}}$. For photons with $\lambda>\lambda_{\rm b}$, they escape from the trapping radius, and could be scattered and absorbed in the outer regions, ultimately being destroyed.\delete {The observed photons  originate from the radius of their last absorption [max($r_{\rm tr}, \, r_{\rm th, \nu}$)].} We analytically compute $\lambda_{\rm b}$ and the monochromatic luminosity $\lambda L_{\lambda}$ in the NIR band as in Eqs. \eqref{spec_num} and \eqref{lambda_b}, which are sensitive to the outflow parameters ($\dot{M}, \, v_{\rm out}$). This enables us to estimate $\dot{M}$ and $v_{\rm out}$ from the observed SED.

In our work,  we obtain the outflow parameters ($\dot{M}, \, v_{\rm out}$) by fitting the outflow reprocessing model to the observed SED of AT2018cow. By integrating the mass loss rate $\dot{M}$ {over time}, we estimate that the total mass of the outflow in AT2018cow {as} $M_{\rm out} \approx 5.7_{-0.4}^{+0.4} \, M_{\odot}$. In Section \ref{Introduction}, we argued that AT2018cow is likely a massive stellar explosion event. For massive stars like BSGs, the mass before the stellar explosion is generally lager than $20 \, M_{\odot}$ \citep[]{Woosley2012}. Therefore, {by the subtraction} we estimate the mass of the central remnant in AT2018cow to be $M_{\rm obj} \gtrsim 14 \, M_{\odot}$. {This implies that the central object in AT2018cow is likely a stellar-mass BH.}

Our conclusion reveals that the central engine of AT2018cow is very likely to be {a BH} accretion disk. When a massive star explodes, the core collapses to form a stellar-mass {BH}, while the outer envelope falls back and forms an accretion disk due to its sufficient specific angular momentum. The high levels of optical polarization observed 12.9 days after the explosion also suggests that the central engine is likely an accretion disk in AT2018cow \citep[]{Maund2023}, which supports this scenario.

\section{ACKNOWLEDGMENTS}
We thank the anonymous reviewer for constructive comments and suggestions that have helped to improve the quality and clarity of the manuscript. This work is supported by National Natural Science Foundation of China (NSFC-12393814 and 12261141691).


\software{Astropy \citep{Price-Whelan2018,Robitaille2013},  emcee \citep[]{emcee2013}.
          }





\clearpage

\bibliography{sample631}{}
\bibliographystyle{aasjournal}


\end{CJK*}
\end{document}